\documentclass[draftcls, 10pt, onecolumn]{IEEEtran}
\usepackage{amssymb, amsmath, graphicx, paralist,subfigure}
\usepackage{amsbsy}
\usepackage{floatflt} 

\usepackage{amsmath}
\usepackage{amssymb}
\usepackage{times}
\usepackage{graphicx}
\usepackage{xspace}
\usepackage{paralist} 
\usepackage{setspace} 
\usepackage{xypic}
\xyoption{curve}
\usepackage{latexsym}
\usepackage{theorem}
\usepackage{ifthen}
\usepackage{subfigure}

{\theoremheaderfont{\it} \theorembodyfont{\rmfamily}
\newtheorem{lem}{Lemma}
\newtheorem{theorem}{Theorem}

}

\renewcommand{\baselinestretch}{1.5}

\def\ln{{\rm ln}}

\begin{document}
\title{\huge Distributed Sensor Localization in Random Environments using Minimal Number of Anchor Nodes}
\author{Usman~A.~Khan$^\dagger$\thanks{$^\dagger$All authors contributed equally to the paper. This work was partially supported by the DARPA DSO Advanced Computing and Mathematics Program Integrated Sensing and Processing (ISP) Initiative under  ARO grant \#~DAAD 19-02-1-0180, by NSF under grants \#~ECS-0225449 and~\#~CNS-0428404, by ONR under grant \#~MURI-N000140710747, and by an IBM Faculty Award.},~Soummya~Kar$^\dagger$,~and~Jos\'e~M.~F.~Moura$^\dagger$\\
            Department of Electrical and Computer Engineering\\
            Carnegie Mellon University, 5000 Forbes Ave, Pittsburgh, PA 15213\\
            \{ukhan, moura\}@ece.cmu.edu,~soummyak@andrew.cmu.edu\\
            Ph: (412)268-7103 Fax: (412)268-3890
}

\maketitle
\begin{abstract}
The paper develops DILOC, a \emph{distributive}, \emph{iterative} algorithm that locates~$M$ sensors in $\mathbb{R}^m, m\geq 1$, with respect to a minimal number of~$m+1$ anchors with known locations. The sensors exchange data with their neighbors only; no centralized data processing or communication occurs, nor is there centralized knowledge about the sensors' locations. DILOC uses the barycentric coordinates of a sensor with respect to its neighbors that are computed using the Cayley-Menger determinants. These are the determinants of matrices of inter-sensor distances. We show convergence of DILOC by associating with it an absorbing Markov chain whose absorbing states are the anchors. We introduce a stochastic approximation version extending DILOC to random environments when the knowledge about the intercommunications among sensors and the inter-sensor distances are  noisy, and the communication links among neighbors fail at random times. We show a.s.~convergence of the modified DILOC and characterize the error between the final estimates and the true values of the sensors' locations. Numerical studies illustrate DILOC under a variety of deterministic and random operating conditions.
\end{abstract}


\newpage
\section{Introduction}
Localization is a fundamental problem in sensor networks.  Information about the location of the sensors is key to process the sensors' measurements accurately. In applications where sensors are deployed randomly, they have no knowledge of their exact locations, but equipping each of them with a localization device like a GPS is expensive, not robust to jamming in military applications, and is usually of limited use in indoor environments. Our goal is to develop a distributed (decentralized) localization algorithm where the sensors find their locations under a limited set of assumptions and conditions. To be more specific, we are motivated by applications where~$N=M+m+1$ sensors in~$\mathbb{R}^m$ (for example, $m=2$ corresponds to sensors lying on a plane, while for $m=3$ the sensors are in three dimensional Euclidean space) are deployed in a large geographical region. We assume that the deployment region lies in the convex hull of a small, in fact minimal, number of~$m+1$ anchors, $(m+1)\ll M$. The anchors know their locations. In such situations, the large geographical distances to the anchors makes it highly impractical for the $M$~non-anchor sensors to communicate directly with the anchors. Further, to compute the locations of the non-anchor sensors at a central station is not feasible when~$M$ is large, as it requires large communication effort, expensive large-scale computation, and adds latency and bottlenecks to the network operation. These networks call for efficient \emph{distributed} algorithms where each sensor communicates directly only with a few neighboring nodes (either sensors or anchors) and a low order computation is performed locally at the sensor and at each iteration of the algorithm, for example, see \cite{usman_tsp:07}. In this paper, we present a Distributed Iterative LOCalization algorithm (DILOC, pronounced {\it die}-{\it lock}) that overcomes the above challenges in large-scale randomly deployed networks.

In DILOC, the sensors start with an initial estimate of their locations, for example, a random guess. This random guess is arbitrary and does not need to place the sensors in the convex hull of the anchors. The sensors then update their locations, which we call the state of the network, by exchanging their state information \emph{only} with a carefully chosen subset of~$m+1$ of their neighbors. This state updating is a convex combination of the states of the neighboring nodes. The coefficients of the convex combination are the barycentric coordinates \cite{riemann_book,top_book} and are determined from the mutual inter-sensor distances among the sensors using the Cayley-Menger determinants. At each sensor~$l$ its neighborhood set contains~$m+1$ sensors, for example, its closest~$m+1$ sensors, such that sensor~$l$ lies in the convex hull of these~$m+1$ neighbors. These neighbors may or may not include the anchors.

DILOC is distributed and iterative; each sensor updates locally its own state and then sends its state information to its neighbors; nowhere does DILOC need a fusion center or global communication. We prove almost sure~(a.s.) convergence of DILOC in both deterministic and random network environments by showing that DILOC behaves as an absorbing Markov chain, where the anchors are the absorbing states. We prove convergence under a broad characterization of noise. In particular, we consider three types of randomness, acting simultaneously. These model many practical random sensing and communication distortions as, for example, when: \begin{inparaenum}[(i)] \item the inter-sensor distances are known up to random errors, which is common in cluttered environments and also in ad-hoc environments, where cheap low resolution sensors are deployed; \item the communication links between the sensors fail at random times. This is mainly motivated by wireless digital communication, where packets may get dropped randomly at each iteration, particularly, if the sensors are power limited or there are bandwidth or rate communication constraints in the network; and \item the communication among two sensors, when their communication link is active, is corrupted by noise.
\end{inparaenum}

Although a sensor can only communicate directly with its neighbors (e.g., sensors within a small radius), we assume that, when the links are deterministic and never fail,  the network graph is connected, i.e., there is a communication path (by multihop) between any arbitrary pair of sensors.
 In a random environment, inter-sensor communication links may not stay active all the time and are subject to random failures. Consequently, there may be iterations when the network is not connected; actually, there might never be iterations when the network is connected. We will show under broad
 conditions almost sure convergence of an extended version of DILOC that we term as Distributed Localization in Random Environments (DLRE). DLRE employs stochastic approximation techniques using a decreasing weight sequence in the iterations. 

 In the following, we contrast our work with the existing literature on sensor localization.

{\bf Brief review of the literature:} The literature on localization algorithms may be broadly
characterized into centralized and distributed algorithms.
Illustrative centralized localization algorithms include: maximum likelihood
estimators that are formulated when the data is known to be described by a statistical model, \cite{patt:03,dea:03};
multi-dimensional scaling~(MDS) algorithms that formulate the
localization problem as a least squares problem at a centralized
location, \cite{fromherz:03,ruml:04}; work that exploits the geometry of the Euclidean space, like when locating a single robot using trilateration in $m=3-$dimensional space, see~\cite{ros:05} where a geometric interpretation is given to the traditional algebraic distance constraint equations;  localization algorithms with imprecise distance information, see~\cite{morse:05} where the authors exploit the geometric relations among the distances in the optimization procedure; for additional work, see, e.g.,
\cite{saul:00,hero:04}. Centralized algorithms are fine in small or tethered network environments; but in large untethered networks, they incur high communication cost and may not be scalable; they depend
on the availability and robustness of a central processor and have a single point of failure.

Distributed localization algorithms can be characterized into two classes: multilateration and successive refinements. In multilateration algorithms, \cite{nath:01,sriv:01,sriv:02,bach:03}, each sensor estimates its range from the anchors and then calculates its location via multilateration, \cite{caff:99}. The multilateration scheme requires a high density of anchors, which is a practical limitation in large sensor networks. Further, the location estimates obtained from multilateration schemes are subject to large errors because the estimated sensor-anchor distance in large networks, where the anchors are far apart, is noisy. To overcome this problem, a high density of anchors is required. We, on the other hand, do not estimate distances to far-away nodes. Only local distances to nearby nodes are estimated that have better accuracy. This allows us to employ a minimal number of anchors. 

A distributed multidimensional scaling algorithm is presented in \cite{hero_dist:03}. Successive refinement algorithms that perform an iterative minimization of a cost function are presented in \cite{zhang:01,beut:01,hubaux:01}. Reference \cite{zhang:01} discusses an iterative scheme where they assume~$5\%$ of the nodes as anchors. Reference \cite{hubaux:01} discusses a Self-Positioning Algorithm (SPA) that provides a GPS-free positioning and builds a relative coordinate system.

Another formulation to solve localization problems in a distributed fashion is the probabilistic approach. Nonparametric belief propagation on graphical models is used in~\cite{willsky:04}. Sequential Monte Carlo methods for mobile localization are considered in~\cite{evans:04}. Particle filtering methods have been addressed in~\cite{coates:04} where each sensor stores representative particles for its location that are weighted according to their likelihood. Reference~\cite{thrun:02} tracks and locates mobile robots using such probabilistic methods.

Completion of partially specified distance matrices is considered in \cite{gower1,gower2}. The approach is relevant when the (entire) partially specified distance matrix is available at a central location. The algorithms complete the unspecified distances under the geometrical constraints of the underlying network. The key point to note in our work is that our algorithm is \emph{distributed}. In particular, it does not require a centralized location to perform the computations. 

In comparison with these references, our algorithm, DILOC, is equivalent to solving in a \emph{distributed} and \emph{iterative} fashion a large system of linear algebraic equations where the system matrix is highly sparse. Our method exploits the structure of this matrix, which results from the topology of the communication graph of the network. We prove the a.s.~convergence of the algorithm under broad noise conditions and characterize the bias and mean square error properties of the estimates of the sensor locations obtained by DILOC.

We divide the rest of the paper into two parts. The first part of the paper is concerned with the deterministic formulation of the localization problem and consists of sections~\ref{PF}--\ref{rel}. Section~\ref{PF} presents preliminaries and then DILOC, the distributed iterative localization algorithm, that is based on barycentric coordinates, generalized volumes, and Cayley-Menger determinants. Section~\ref{senm} proves DILOC's convergence. Section~\ref{rel} presents the DILOC-REL, DILOC with relaxation, and proves that it asymptotically reduces to the deterministic case without relaxation. The second part of the paper consists of sections~\ref{AlgAss}--\ref{asconv} and considers distributed localization in random noisy environments. Section~\ref{AlgAss} characterizes the random noisy environments and the iterative algorithm for these conditions. Section~\ref{asconv} proves the convergence of the distributed localization algorithm in the noisy case that relies on a result on the convergence of Markov processes. Finally, we present detailed numerical simulations in Section~\ref{num} and conclude the paper in Section~\ref{conc}. Appendices~\ref{CHIT}--\ref{IR} provide a necessary test, the Cayley-Menger determinant and background material on absorbing Markov chains.

\section{Distributed Sensor Localization: DILOC}\label{PF}
In this section, we formally state DILOC (distributed iterative localization algorithm) in $m$-dimension Euclidean space, $\mathbb{R}^m~(m\geq 1)$, and introduce the necessary notation. Of course, for sensor localization,  $m=1$ (sensors in a straight line), $m=2$ (plane), or~$m=3$ ($3$d-space.) The generic case of~$m>3$ is of interest, for example, when the graph nodes represent $m$-dimensional feature vectors in classification problems, and the goal is still to find in a distributed fashion their global coordinates (with respect to a reference frame.) Since our results are general, we deal with $m$-dimensional `localization,' but, for easier reference, the reader may consider~$m=2$ or~$m=3$. To provide a quantitative assessment on some of the assumptions underlying DILOC, we will, when needed, assume that the deployment of the sensors in a given region follows a Poisson distribution. This random deployment is often assumed and is realistic; we use it to derive probabilistic bounds on the deployment density of the sensors and on the communication radius at each sensor; these can be straight forwardly related to the values of network field parameters (like transmitting power or signal-to-noise ratio) in order to implement DILOC. We discuss the computation/communication complexity of the algorithm and provide a simplistic, yet insightful, example that illustrates DILOC.

\subsection{Preliminaries and Notation}
Recall that the sensors are in $\mathbb{R}^m$. Let $\Theta$ be the set of sensors or nodes in the network decomposed as
\begin{equation}
\Theta=\kappa\cup\Omega,
\end{equation}
where~$\kappa$ is the set of anchors, i.e., the sensors whose locations are known, and~$\Omega$ is the set of sensors whose locations are to be determined. By $|\cdot|$ we mean the cardinality of the set, and we let $\left|\Theta\right|=N$, $\left|\kappa\right|=m+1$, and $\left|\Omega\right|=M$.
For a set~$\Psi$ of sensors, we denote its convex hull by $\mathcal{C}\left(\Psi\right)$\footnote{The convex hull, $\mathcal{C}\left(\Psi\right)$, of a set of points in~$\Psi$ is the minimal convex set containing~$\Psi$.}. For example, if $\Psi$ is a set of three non-coplanar sensors in a plane, then $\mathcal{C}\left(\Psi\right)$ is a triangle. We now define a few additional sets needed.

Let $d_{lk}$ be the Euclidean distance between two sensors $l,k\in\Theta$. We associate with the sensor~$l\in\Omega$, a positive real number,~$r_l>0$, and two sets $\mathcal{K}\left(l,r_l\right)$ and~$\Theta_l\left(r_l\right)$:
\begin{eqnarray}
\label{ne_l}
&\mathcal{K}\left(l,r_l\right)=\left\{k\in\Theta:\:\:d_{lk}<r_l\right\},&\\
\label{linC}
&\Theta_l\left(r_l\right)\subseteq\mathcal{K}\left(l,r_l\right),\:\: l\notin\Theta_l\left(r_l\right),\:\:
l\in\mathcal{C}\left(\Theta_l\left(r_l\right)\right),\:\:
\left|\Theta_l\left(r_l\right)\right|=m+1,\:\:
A_{\Theta_l\left(r_l\right)}\neq 0,&
\end{eqnarray}
where $A_{\Theta_l\left(r_l\right)}$ is the generalized volume (area in $m=2$-d, volume in $m=3$, and their generalization in higher dimensions) of~$\mathcal{C}\left(\Theta_l\left(r_l\right)\right)$. The set $\mathcal{K}\left(l,r_l\right)$ groups the neighboring sensors of~$l$ within a radius~$r_l$ and, by~(\ref{linC}), $\Theta_l\left(r_l\right)$, which we will often represent simply as~$\Theta_l$, assuming $r_l$ is understood from the context, is a subset of~$m+1$ sensors such that sensor~$l$ lies in its convex hull but is not one of its elements. In appendix~\ref{CHIT}, we provide a procedure to test the convex hull inclusion of a sensor, i.e., for any sensor, $l$, to determine if it lies in the convex hull of~$m+1$ nodes arbitrarily chosen from the set, $\mathcal{K}\left(l,r_l\right)$, of its neighbors. Finding such a set~$\Theta_l$ is an important step in DILOC and we refer to it as \emph{triangulation} and $\Theta_l$ is referred to as a \emph{triangulation set}.

Let~$\mathbf{c}_l$ be the $m$-dimensional coordinates for a node, $l\in\Theta$, with respect to a global coordinate system, written as the $m$-dimensional row vector,
\begin{eqnarray}
\label{eq_coord}
\mathbf{c}_l &=& \left[c_{l,1},c_{l,2},\ldots,c_{l,m}\right].
\end{eqnarray}
The true (possibly unknown) location of sensor~$l$ is represented by $\mathbf{c}_l^\ast$. Because the distributed localization algorithm DILOC is iterative, $\mathbf{c}_l(t)$ will represent the location vector, or state, for sensor~$l$ at iteration~$t$.

\textbf{Barycentric coordinates.} DILOC is expressed in terms of the barycentric coordinates,~$a_{lk}$, of a sensor,~$l\in\Omega$, with respect to the nodes, $k\in\Theta_l$, see~\cite{riemann_book,top_book}. The barycentric coordinates,~$a_{lk}$, are unique and are given by
\begin{equation}
\label{alk}
a_{lk}=\dfrac{A_{\{l\}\cup \Theta_l  \setminus \{k\}}}{A_{\Theta_l}},
\end{equation}
with $A_{\Theta_l}\neq0$, where `$\setminus$' denotes the set difference, $A_{\{l\}\cup \Theta_l \setminus\{k\}}$ is the generalized volume of the set $\{l\}\cup \Theta_l\setminus\{k\}$, i.e., the set~$\Theta_l$ with sensor~$l$ added and node~$k$ removed.  The barycentric coordinates can be computed from the inter-sensor distances $d_{lk}$ using the Cayley-Menger determinants as shown in appendix~\ref{CMdet}. From~\eqref{alk}, and the facts that the generalized volumes are non-negative and
\begin{equation}
\label{eq:generalizedvolume-a}
\sum_{k\in\Theta_l} A_{\Theta_l\cup \{l\}\setminus\{k\}} = A_{\Theta_l},\qquad l\in\mathcal{C}(\Theta_l),
\end{equation}
it follows that, for each $l\in\Omega$, $k\in\Theta_l$,
\begin{equation}
\label{eq:aklsumone}
a_{lk}\in[0,~1],\qquad
\sum_{k\in\Theta_l} a_{lk} = 1.
\end{equation}

\subsection{Distributed iterative localization algorithm.}
\label{diloc_sec}
Before presenting DILOC, we state explicitly state our assumptions.

\textbf{(B0)~Nondegeneracy.}  The generalized volume for~$\kappa$, $A_\kappa\neq0$.\footnote{Nondegeneracy simply states that the anchors do not lie on a hyperplane. If this was the case, then the localization problem reduces to a lower dimensional problem, i.e., $\mathbb{R}^{m-1}$ instead of $\mathbb{R}^m$. For instance, if all the anchors in a network lie on a plane in~$\mathbb{R}^3$, the localization problem can be thought of as localization in~$\mathbb{R}^2$.}

\textbf{(B1)~Anchor nodes.} The anchors' locations are known, i.e., their state remains constant
\begin{equation}
\label{eq:anchorknown}
\mathbf{c}_q(t) =\mathbf{c}_q^\ast,\qquad q\in\kappa, \: t\geq0.
\end{equation}

\textbf{(B2)~Convexity.} All the sensors lie inside the convex hull of the anchors
\begin{equation}
\label{kinO}
\mathcal{C}(\Omega)\subset\mathcal{C}(\kappa).
\end{equation}

From~(B2), the next Lemma follows easily.
\begin{lem}[Triangulation]
\label{lemma:triangulation1}
For every~$l\in\Omega$, there \emph{exist} $r_l>0$ and~$\Theta_l\left(r_l\right)$ with $\left|\Theta_l\left(r_l\right)\right|=m+1$ satisfying the properties in~\eqref{linC}.
\end{lem}
\begin{proof}
Clearly, by~(B2), $\kappa$ satisfies~\eqref{linC} and the diameter of the network, $\max_{l,k} d_{lk}, (l\in\Omega,k\in\kappa )$, could be taken as~$r_l$. \end{proof} Lemma~\ref{lemma:triangulation1} provides an existence proof, but in localization in wireless sensor networks, it is important to triangulate a sensor not with the network diameter but with a small~$r_l$. In fact, Section~\ref{rand_dep} discusses the probability of finding one such $\Theta_l$ with $r_l\ll\max_{l,k} d_{lk}, (l\in\Omega,k\in\kappa )$.

As a note, it is easily verified that every pair of sensors
in~$\mathcal{C}\left(\Theta_l\left(r_l\right)\right)$ is within a distance $R_l=2r_l$. Think of~$m=2$, then $\left|\Theta_l\left(r_l\right)\right|=3$, $\mathcal{C}\left(\Theta_l\left(r_l\right)\right)$ is a triangle and if $r_l$ is the maximum distance of any interior point~$l$ to the vertices of the triangle, the maximum distance between the 4~points (triangle vertices and~$l$) is $2r_l$. We complete stating the assumptions underlying DILOC.

\textbf{(B3)~inter-sensor distances.} For $l\in\Omega$, there exists at least an $r_l>0$ and $\Theta_l\left(r_l\right)\subset \mathcal{K}\left(l, r_l\right)$, satisfying~\eqref{linC}, such that~$l$  has a communication link for every $k\in \Theta_l\left(r_l\right)$ and knows the mutual distances among all the nodes in~$\{l\}\cup\Theta_l\left(r_l\right)$.

We can now present DILOC. There are two steps: a set-up phase and then DILOC proper. We discuss each separately.

\textbf{DILOC set-up: Triangulation.} In the set-up phase, each sensor~$l$ triangulates itself, so that by the end of this phase we have paired every~$l\in\Omega$ with its corresponding~$m+1$ neighbors in $\Theta_l$. Since triangulation should be with a small~$r_l$, the following is a practical protocol for the set-up phase. A sensor starts with a small communication radius, $r_l$, and chooses arbitrarily~$m+1$ sensors within $r_l$ and tests if it lies in the convex hull of these sensors using the procedure described in Appendix~\ref{CHIT}. The sensor attempts this with all collections of~$m+1$ sensors within~$r_l$. If all attempts fail, the sensor adaptively increases, in small increments, its communication radius, $r_l$, and repeats the process. By~(B2) and~\eqref{kinO}, success is eventually achieved and each sensor is triangulated by finding~$\Theta_l$ with properties~\eqref{linC} and~(B3). If the sensors have directionality a much simpler algorithm based on Lemma~\ref{lem_sc1} below, see also discussion following the Lemma, triangulates the sensor with high probability of success in one shot. To assess the practical implications required by DILOC's set-up phase, Subsection~\ref{rand_dep} considers the realistic scenario where the sensors are deployed using a random Poisson distribution and computes in terms of deployment parameters the probability of finding at least one such~$\Theta_l$ in a given radius, $r_l$.

\textbf{DILOC iterations: state updating.} Once the set-up phase is complete; at time~$t+1$, each sensor~$l\in\Omega$, iteratively updates its state, i.e., its current location estimate, by a convex combination of the states at time~$t$ of the nodes in~$\Theta_l$. The anchors do not update their state, since they know their locations. The updating is explicitly given by
\begin{equation}
\label{diloc}
\mathbf{c}_l(t+1) = \left\{\begin{array}{cc}\mathbf{c}_l(t),&l\in\kappa,\\
\sum_{k\in\Theta_l}a_{lk}\mathbf{c}_k(t),&l\in\Omega,
\end{array}\right.
\end{equation}
where $a_{lk}$ are the barycentric coordinates of~$l$ with respect to $k\in\Theta_l$.
DILOC in~\eqref{diloc} is distributed since (i) the update is implemented at each sensor independently; (ii) at sensor $l\in\Omega$, the update of the state,~$\mathbf{c}_l(t+1)$, is obtained from the states of its~$m+1$ neighboring nodes in~$\Theta_l$; and (iii) there is no central location and only local information is available.

\textbf{DILOC: Matrix format.} For compactness of notation, we write DILOC~\eqref{diloc} in matrix form. Without loss of generality, we index the anchors in $\kappa$ as $1,2,\ldots,m+1$ and the sensors in $\Omega$ as $m+2,m+3,\ldots,m+1+M=N$.  We now stack the (row vectors) states,~$\mathbf{c}_l$, given in~\eqref{eq_coord} for all the~$N$ nodes in the network in the~$N\times m$-dimensional coordinate matrix
\begin{equation}
\label{C_coord}
\mathbf{C}=\left[\begin{array}{c}\mathbf{c}_1^T,\ldots, \mathbf{c}_N^T\end{array}\right]^T.
\end{equation}
DILOC equations in~\eqref{diloc}, now become in compact matrix form
\begin{equation}
\label{diloc_mat}
\mathbf{C}(t+1) = \mathbf{\mathbf{\Upsilon}}\mathbf{C}(t).
\end{equation}
The structure of the $N\times N$ iteration matrix~$\mathbf{\Upsilon}$ becomes more apparent if we partition it as
 \begin{equation}
 \label{Up_part}
 \mathbf{\Upsilon}=\left[\begin{array}{cc}
\mathbf{I}_{m+1}&\mathbf{0}\\
\mathbf{B}&\mathbf{P}
\end{array} \right],
\end{equation}
The first $m+1$ rows correspond to the update equations for the anchors in~$\kappa$. Since the states of the anchors are constant, see~(B1) and~\eqref{eq:anchorknown}, the first $m+1$ rows of $\mathbf{\Upsilon}$ are zero except for a~$1$ at their $(q,q), q\in\kappa=\{1,\ldots,m+1\}$ location. In other words, these first~$m+1$ rows are the $(m+1)\times N$ block matrix $\left[\mathbf{I}_{m+1}|\mathbf{0}\right]$, i.e., the $(m+1)\times (m+1)$ identity matrix $\mathbf{I}_{m+1}$ concatenated with the $(m+1)\times M$ zero matrix,~$\mathbf{0}$.

Each of the $M$ remaining rows in~$\mathbf{\Upsilon}$, indexed by $l\in\Omega=\{m+2,m+3,\ldots,N\}$, have only $m+1$ non-zero elements corresponding to the nodes in the triangulation set,~$\Theta_l$, of~$l$, and these non-zero elements are the barycentric coordinates, $a_{lk}$, of sensor~$l$ with respect to the nodes in $\Theta_l$. The $M\times (m+1)$ block $\mathbf{B}=\{b_{lj}\}$ is a zero matrix, except in those rows corresponding to the sensors in~$\Omega$ that have a direct link to anchors. The $M\times M$  block $\mathbf{P}=\left\{p_{lj}\right\}$  is also a sparse matrix where the non-zero entries in row~$l$ correspond to the non-anchor nodes in~$\Theta_l$. The matrices~$\mathbf{\Upsilon}$, $\mathbf{P}$, and~$\mathbf{B}$ have important properties that will be used to prove the convergence of the distributed iterative algorithm DILOC in Sections~\ref{senm} and~\ref{rel}.

\textbf{Remark.} Equation~\eqref{diloc_mat} writes DILOC in matrix format for compactness; it should not be confused with a centralized algorithm--it still is a \emph{distributed} iterative algorithm. It is iterative, because each iteration through~\eqref{diloc_mat} simply updates the (matrix of) state(s) from~$\mathbf{C}(t)$ to~$\mathbf{C}(t+1)$. It is distributed because each row equation updates the state of sensor~$l$ from the states of the~$m+1$ nodes in~$\Theta_l$. In all, the iteration matrix, $\mathbf{\Upsilon}$, is highly sparse having exactly $(m+1)+ M(m+1)$ non-zeros out of possible $(m+1+M)^2$ elements.
\subsection{Example}\label{example}
\begin{figure}
\centering
\includegraphics[width=2in]{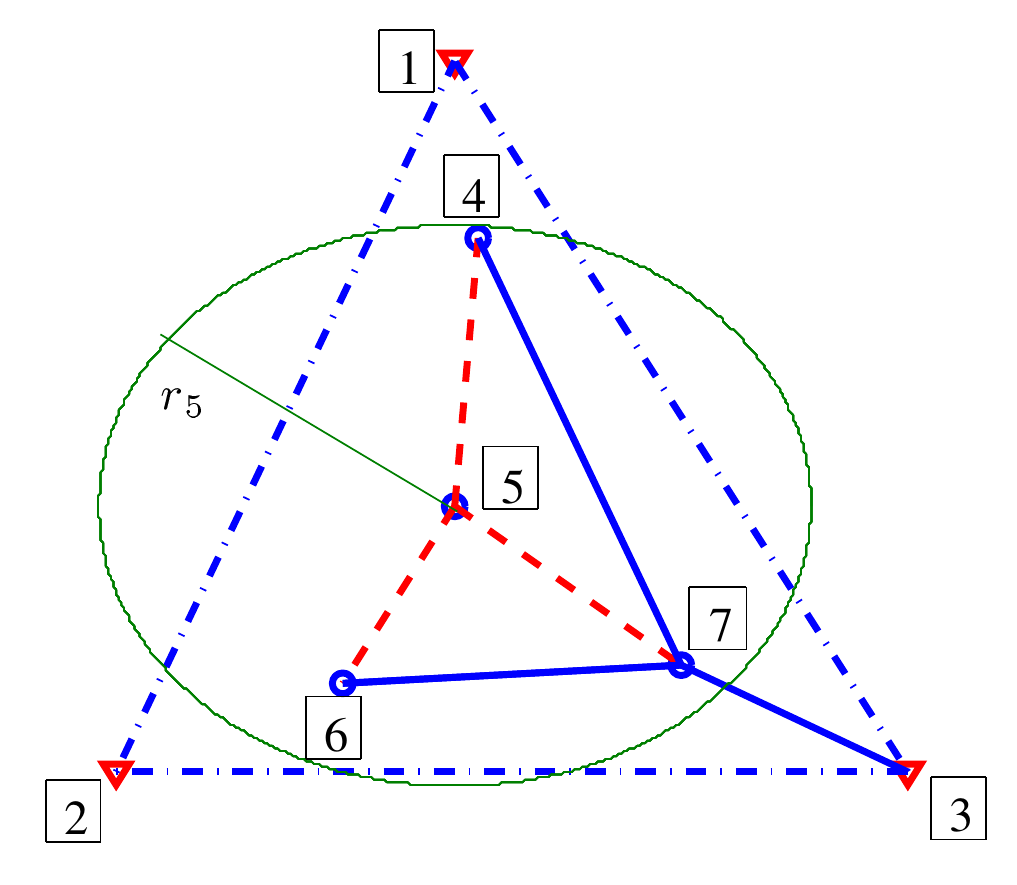}
\caption{Deployment corresponding to the example in Section~\ref{example}.}
\label{examp_network}
\end{figure}

We consider an $m=2$-dimensional Euclidean plane with $m+1=3$ anchors and~$M=4$ sensors, see Fig.~\ref{examp_network}.
The nodes are indexed such that the anchor set is $\kappa=\{1,2,3\}, ~|\kappa|=m+1=3,$ and the sensor set is $\Omega=\{4,5,6,7\}, \,|\Omega|=M=4.$ The set of all the nodes in the network is, thus, $\Theta=\kappa\cup\Omega=\{1,\ldots,7\},~|\Theta|=N=7$. The triangulation set, $\Theta_l,~l\in\Omega$, identified by using the convex hull inclusion test are $\Theta_4=\{1,5,7\},~\Theta_5=\{4,6,7\},~\Theta_6=\{2,5,7\},~\Theta_7=\{3,4,6\}.$ It is clear that these triangulation sets satisfy properties in~\eqref{linC}. It can be verified that sensor $5$ does not have any anchor node in its triangulation set, $\Theta_5$, and every other sensor has only one anchor in its respective triangulation set. Since, no sensor is able to communicate to $m+1=3$ anchors directly, no sensor can localize itself in a single step.

At each sensor, $l\in\Omega$, the barycentric coordinates, $a_{lk},~k\in\Theta_l$, are computed using the inter-sensor distances (among the nodes in the set $\{l\}\cup\Theta_l$) in the Cayley-Menger determinant. It is noteworthy that the inter-sensor distances that need to be known at each sensor~$l$ to compute $a_{lk}$ are only the inter-sensor distances among the $m+2$~sensors in the set $\{l\}\cup\Theta_l$. For instance, the distances in the Cayley-Menger determinant needed by sensor~$5$ to compute $a_{54},a_{56},a_{57}$ are the inter-sensor distances among the nodes in the set, $\{5\}\cup\Theta_5$, i.e., $d_{54},d_{56},d_{57},d_{46},d_{47},d_{67}$. Due to~\eqref{linC} and the relation $\Theta_5\subseteq\mathcal{K}(l,r_5)$, the nodes in $\{5\}\cup\Theta_5$ lie in a circle of radius, $r_5$, centered at sensor $5$ (shown in Fig.~\ref{examp_network}); on the other hand, no two sensors in the set $\{5\}\cup\Theta_5$ can be more than $R_5=2r_5$ apart. This justifies the choices of~$R_l$ and  $r_l=R_l/2$ in~\eqref{ne_l}.

Once the barycentric coordinates, $a_{lk}$, are computed, DILOC for the sensors in $\Omega$ is \begin{eqnarray}\label{ex_eq}\mathbf{c}_l(t+1)&=&\sum_{k\in\Theta_l}a_{lk}\mathbf{c}_k(t),\qquad l\in\Omega=\{4,5,6,7\}.\end{eqnarray} In particular, for sensor $5$, we have the following expression,
$$\mathbf{c}_5(t+1) = a_{54}\mathbf{c}_4(t)+a_{56}\mathbf{c}_6(t)+a_{57}\mathbf{c}_7(t).$$
DILOC for the anchors is given by
\[
\mathbf{c}_q(t)=\mathbf{c}_q^\ast,\qquad q\in\kappa.
\]
We write DILOC for this example in the matrix format~\eqref{diloc_mat}.
\begin{eqnarray}\label{diloc_mat_ex}
\left[\begin{array}{c}\mathbf{c}_1(t+1)\\\mathbf{c}_2(t+1)\\\mathbf{c}_3(t+1)\\\mathbf{c}_4(t+1)\\\mathbf{c}_5(t+1)\\\mathbf{c}_6(t+1)\\ \mathbf{c}_7(t+1)\\\end{array}\right]=\left[\begin{array}{ccccccc}
1&0&0&0&0&0&0\\
0&1&0&0&0&0&0\\
0&0&1&0&0&0&0\\
a_{41}&0&0&0&a_{45}&0&a_{47}\\
0&0&0&a_{54}&0&a_{56}&a_{57}\\
0&a_{62}&0&0&a_{65}&0&a_{67}\\
0&0&a_{73}&a_{74}&0&a_{76}&0\\
\end{array}\right]\left[\begin{array}{c}\mathbf{c}_1(t)\\\mathbf{c}_2(t)\\\mathbf{c}_3(t)\\ \mathbf{c}_4(t)\\\mathbf{c}_5(t)\\\mathbf{c}_6(t)\\ \mathbf{c}_7(t)\\\end{array}\right],
\end{eqnarray} where the initial condition are $\mathbf{C}(0)=[\mathbf{c}_1^\ast,\mathbf{c}_2^\ast,\mathbf{c}_3^\ast, \mathbf{c}_4^0, \mathbf{c}_5^0, \mathbf{c}_6^0, \mathbf{c}_7^0]^T$, with $\mathbf{c}_l^0,~l\in\Omega,$ being randomly chosen row vectors of appropriate dimensions.

Note here again that~\eqref{diloc_mat_ex} is just a matrix representation of~\eqref{ex_eq}. DILOC is implemented in a distributed fashion as in~\eqref{ex_eq}. The matrix representation in~\eqref{diloc_mat_ex} is included for compaction of notation and for the convergence analysis of the algorithm.
%
%
%
%
%
%
\subsection{Complexity of DILOC}
Once the barycentric coordinates are computed, each sensor\footnote{By sensor, we usually mean a non-anchor node.} performs the update in~\eqref{diloc} that requires~$m+1$ multiplications and~$m$ additions. Assuming the computation complexity of the multiplication and the addition operations to be the same, the computation complexity of DILOC is~$2m+1$ operations, i.e., $O(1)$ per sensor, per iteration. Since each sensor exchanges information with~$m+1$ nodes in its triangulation set, the communication complexity of DILOC is~$m+1$ communications, i.e., $O(1)$ per sensor, per iteration. Hence, both the computation and communication complexity are $O(M)$ for a network of $M$ sensors. Note that, since the triangulation set-up phase, which identifies $\Theta_l\left(r_l\right)$ and computes the barycentric coordinates, as explained in Subsection~\ref{diloc_sec}, are to be carried out only once at the start of DILOC, they require a constant computation/communication complexity, so we do not account explicitly for it.
\subsection{Random Poisson Deployment}
\label{rand_dep}
A common model in wireless sensor networks is  the Poisson deployment \cite{hall_book,swami:05}. We illustrate it on the plane, $m=2$; the discussion can be extended to arbitrary dimensions. For a Poisson distribution with density, $\gamma >0$, the mean number of sensors in a sector~$Q$ with area~$A_Q$ is $\gamma A_Q$.  The numbers of sensors in any two disjoint sectors, $Q_1$ and~$Q_2$, are independent random variables, and the locations of the sensors in a sector~$Q$ are uniformly distributed. We now characterize the probability of triangulating a sensor~$l$ in a region of radius, $r_l$, centered at~$l$.
\begin{figure}[htp]
\centering
\subfigure[]{
\includegraphics[width=2in]{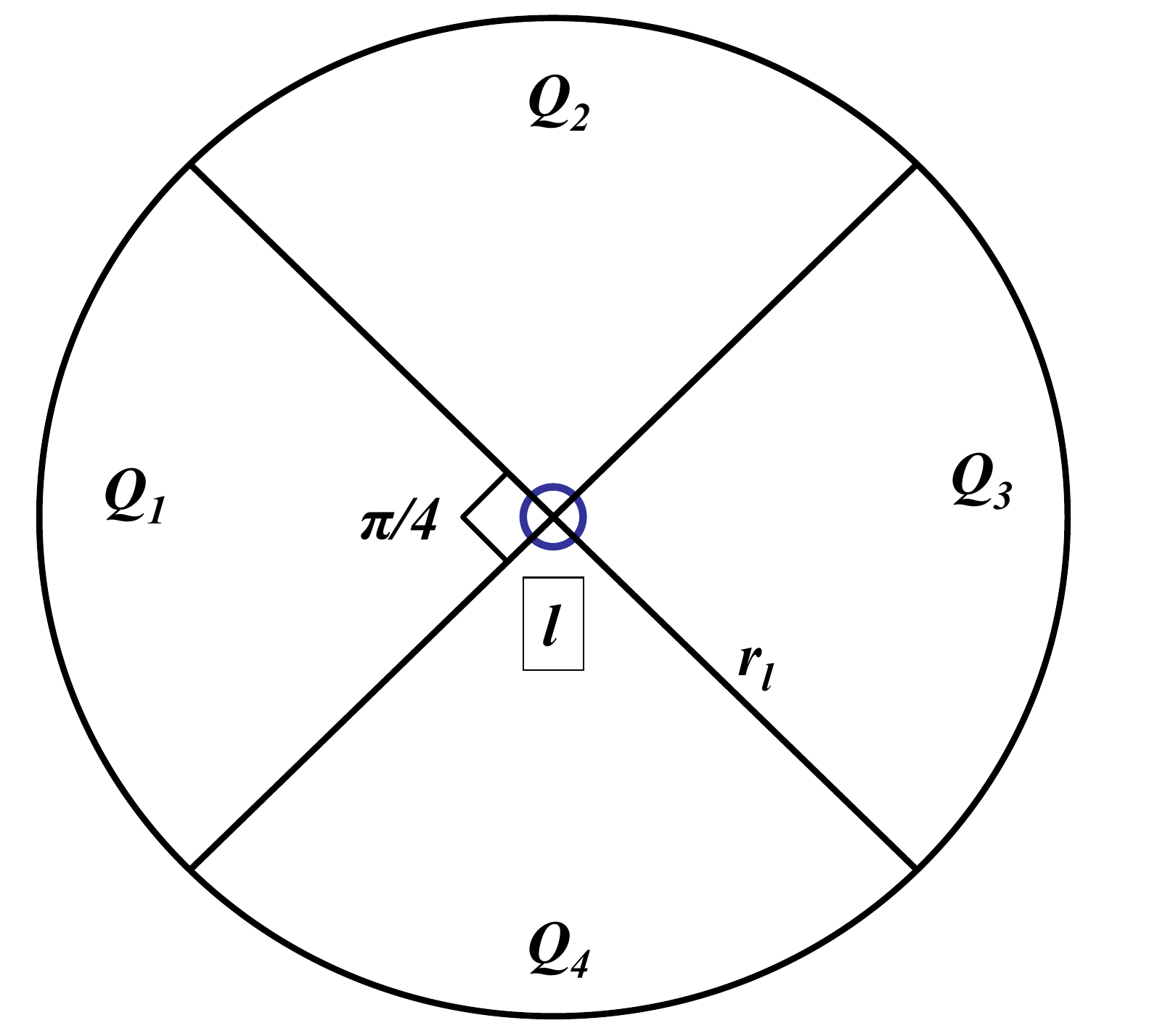}
\label{dep_fig}
}
\hspace{1cm}
\subfigure[]{
\includegraphics[width=2in]{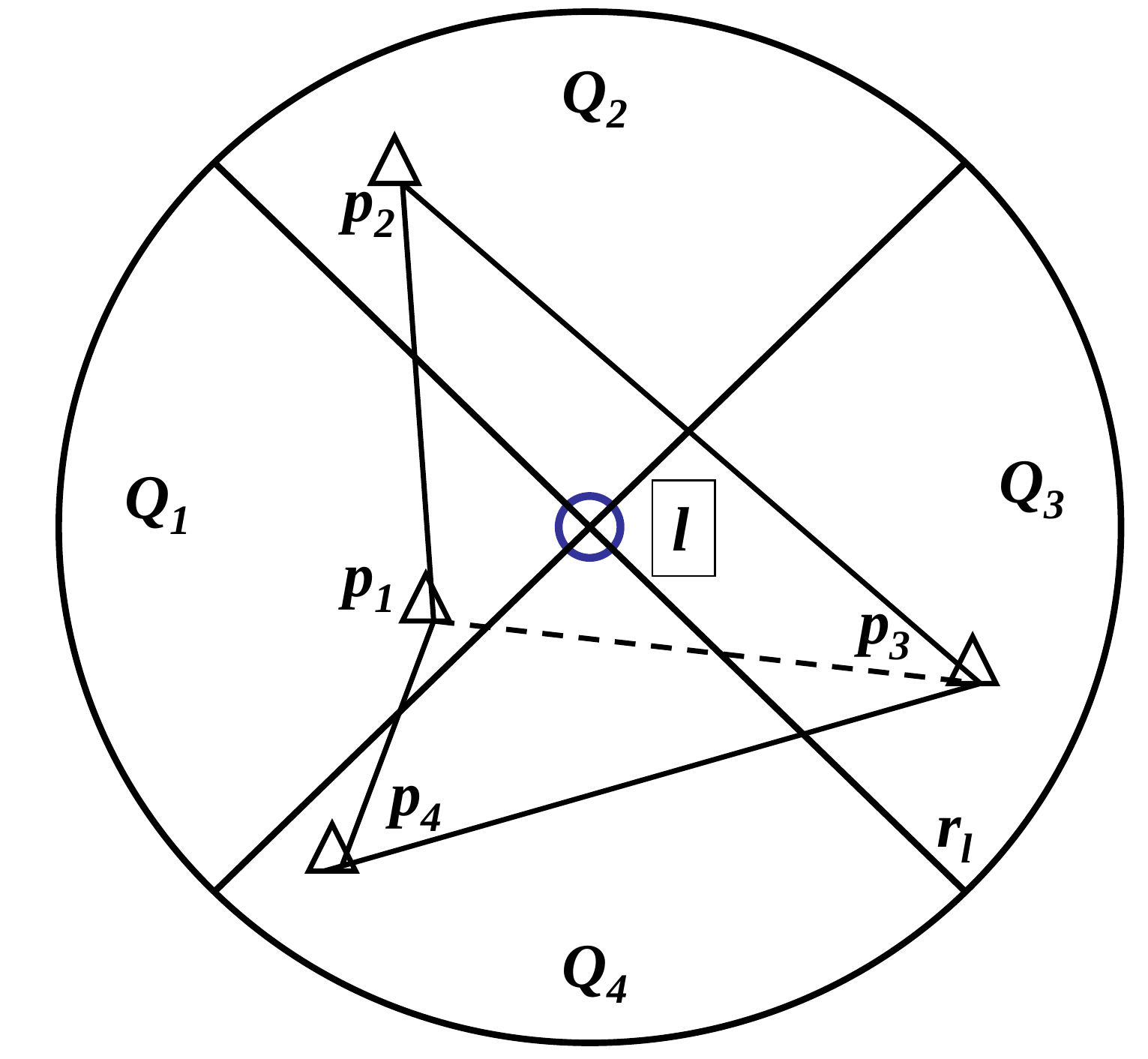}
\label{lem_sc}
}
\caption{(a) Sensor~$l$ identifies its triangulation set, $\Theta_l$, in the circle of radius, $r_l$, centered at sensor $l$. The circle is  divided into four disjoint sectors with equal areas, $Q_1,\ldots,Q_4$. A sufficient condition for triangulation is that there exists at least one sensor in each of these four sectors. (b) Illustration of Lemma~\ref{lem_sc1}.}
\end{figure}
To this end, consider Fig.~\ref{dep_fig}, which shows the triangulation region, a circle of radius, $r_l$, centered at sensor $l$. Let $Q_1,Q_2,Q_3,Q_4$ be four disjoint sectors partitioning this circle with equal areas, i.e., $A_{Q_i}=\dfrac{\pi r_l^2}{4},~i=1,\ldots,4$.

\begin{lem}\label{lem_sc1} A sufficient condition to triangulate a sensor $l\in\mathbb{R}^2$ is to have at least one sensor in each of the four disjoint equal area sectors, $Q_i,~i=1,\ldots,4$, which partition the circle of radius of radius,~$r_l$, centered at~$l$.
\end{lem}
\begin{proof} In Fig.~\ref{lem_sc} consider the triangulation of sensor~$l$ located at the center of the circle; we choose arbitrarily four sensors $p_1,p_2,p_3,p_4$ in each of the four sectors $Q_1, Q_2,Q_3,Q_4$. Denote the polygon with vertices $p_1,p_2,p_3,p_4$ by $\mbox{Pol}\left(p_1p_2p_3p_4\right)$. Consider the diagonal\footnote{If $\mbox{Pol}\left(p_1p_2p_3p_4\right)$ is concave, we choose the diagonal that lies inside $\mbox{Pol}\left(p_1p_2p_3p_4\right)$, i.e., $p1$---$p3$ in Fig.~\ref{lem_sc}. If $\mbox{Pol}\left(p_1p_2p_3p_4\right)$ is convex, we can choose any of the two diagonals and the proof follows.} $p1$---$p3$ that partitions this polygon into two triangles $\triangle p_1p_2p_3$ and $\triangle p_1p_3p_4$. Since $l\in\mbox{Pol}\left(p_1p_2p_3p_4\right)$ and $\triangle p_1p_2p_3\cup\triangle p_1p_3p_4=\mbox{Pol}\left(p_1p_2p_3p_4\right)$ with $\triangle p_1p_2p_3\cap\triangle p_1p_3p_4=\varnothing$, then either $l\in\triangle p_1p_2p_3$ or $l\in\triangle p_1p_3p_4$. The triangle in which~$l$ lies becomes the triangulating set, $\Theta_l$, of~$l$.

This completes the proof. The generalization to higher dimensions is straightforward; for instance, in $\mathbb{R}^3$, we have eight sectors and an arbitrary sensor~$l$ is triangulated with at least one sensor in each of these eight sectors (with equal volume) of a sphere of radius, $r_l$, centered around $l$.\end{proof}

Let $\overline{Q}_i$ be the set of sensors in the sector $Q_i$. It follows from the Poisson deployment that the probability of finding at least one sensor in a sector, $Q_i$, of area $A_{Q_i}$ is
\begin{equation}
\mathbb{P}\left(|\overline{Q}_i|>0\right) = \left(1-\exp^{-\gamma A_{Q_i}} \right).
\end{equation}
Since the distribution of the number of sensors in disjoint sectors is independent, the probability of finding at least one sensor in each of the sets, $\overline{Q}_1,\ldots,\overline{Q}_4$, is the product
\begin{equation}
\mathbb{P}\left(\left|\overline{Q}_i\right|>0,\,\forall\,i\right) = \left(1-\exp^{-\gamma \pi r_l^2/4} \right)^4.
\end{equation}
Thus, we have
\begin{equation}
\label{Pth}\mathbb{P}_{\Theta_l}=\mathbb{P}\left(\Theta_l \mbox{ exists satisfying~\eqref{linC}}\right) \geq \mathbb{P}\left(|\overline{Q}_i|>0,~\forall~i\right).
\end{equation}
This shows that, for a given deployment density, $\gamma$, we can choose, $r_l$, appropriately, to guarantee the triangulation with arbitrarily high probability. Indeed, it follows from~\eqref{Pth} that, for an arbitrary, $0<\epsilon<1$,
to get the probability of triangulation to be greater than~$\epsilon$, i.e., $\mathbb{P}_{\Theta_l}\geq\epsilon$, the radius~$r_l$ should be
\begin{equation}
\label{rB}
R_l\geq 2\left(\dfrac{-4 \ln \left(1 - \epsilon^{\frac{1}{4}}\right)}{\gamma \pi} \right)^{\frac{1}{2}},
\end{equation}
In alternative, if we are limited by the communication radius, $R_l$, to guarantee $\mathbb{P}_{\Theta_l}\geq\epsilon$, we will need the deployment density,~$\gamma$, to be larger than
\begin{equation}
 \label{gB}
 \gamma\geq\dfrac{-4}{\pi \left(R_l/2\right)^2}\ln\left(1 - \epsilon^{\frac{1}{4}}\right).
 \end{equation}
For example, if the sensors are deployed (following a Poisson distribution) with a density of $\gamma = 1$ sensor$/$m$^2$, then we can compute from equations~\eqref{rB}--\eqref{gB} that $99$\% of the sensors will be able to triangulate (identify $\Theta_l$) themselves if they can communicate to at least a radius of $R_l=5.52$m. The remaining ($1\%$) of the sensors may require to communicate to a larger radius.

It also follows from the above discussion that if the sensors are equipped with a sense of directionality (for example, if they all have ultrasound transducers) then each sensor has to find one neighbor in each of its $4$~sectors, $Q_{l,1},Q_{l,2},Q_{l,3},Q_{l,4}$ (in $m=2$-d space). Once a neighbor is found, triangulation  chooses~$3$ out of these~$4$, in order to identify $\Theta_l$. The computational complexity in $m=2$-d Euclidean space is $4~\mbox{choose}~3=4$. Without directionality the process of finding $\Theta_l$ has the (expected) computation complexity of $\gamma \pi r_l^2~\mbox{choose}~3$.

\section{Convergence of DILOC}\label{senm}
In this section, we prove the convergence of DILOC to the exact locations of the sensors, $\mathbf{c}_l^\ast,~l\in\Omega$. To formally state the convergence result, we provide briefly some background and additional results, based on assumptions {\bf (B0)}--{\bf (B3)}.

The entries of the rows of the iteration matrix~$\mathbf{\Upsilon}$, in~\eqref{diloc_mat}, are either zero or the barycentric coordinates,~$a_{lk}$, which are positive, and, by~\eqref{eq:aklsumone}, add to 1. This matrix can then be interpreted as the transition matrix of a Markov chain. We then describe localization problem and DILOC in terms of a Markov chain. Let the assumptions {\bf (B0)--(B3)} in Section~\ref{diloc_sec} hold and the~$N$ nodes in the sensor network correspond to the states of a Markov chain where let the ($ij$)-th element of the iteration matrix, $\mathbf{\Upsilon}=\{\upsilon_{ij}\}$ defines the probability that the~$i$th state goes to the~$j$th state. Because of the structure of $\mathbf{\Upsilon}$,  this chain is a very special Markov chain.

\textbf{Absorbing Markov chain.}   Let an $N\times N$ matrix, $\mathbf{\Upsilon}=\{\upsilon_{ij}\}$, denote the transition probability matrix of a Markov chain with $N$ states, $s_{i,i=1,\ldots,N}$. A state $s_i$ is called absorbing if the probability of leaving that state is~$0$ (i.e., $\upsilon_{ij} = 0, i\neq j$, in other words $\upsilon_{ii}=1$). A Markov chain is said to be absorbing if it has at least one absorbing
state, and if from every state it is possible to go with non-zero probability to an absorbing state (not necessarily in one step).  In an absorbing Markov chain, a state which is not absorbing is called transient. For additional background, see, for example, \cite{grinstead_book}.

\begin{lem}
\label{mp_lem}
The underlying Markov chain with the transition probability matrix given by the iteration matrix, $\mathbf{\Upsilon}$, is absorbing.
\end{lem}
\begin{proof} We prove by contradiction. Since $\upsilon_{ii}=1,~i\in\kappa$, the anchors are the absorbing states of the Markov chain. Since $\upsilon_{ii}=0, i\in\Omega$, the (non-anchor) sensors are the transient states. Partition the transient states into two clusters~C1 and~C2, such that each transient state in~C1 can go with non-zero probability to at least one of the absorbing states and, with probability~1, the transient states in~C2 cannot reach an absorbing states. It follows that with probability~1 the transient states in~C2 cannot reach the transient states in~C1 (in one or multiple steps); otherwise, they reach an absorbing state with a non-zero probability. Let's consider the lie on the boundary of the convex hull, $\mathcal{C}($C2$)$, i.e., the vertices of $\mathcal{C}($C2$)$. Because they are on boundary, they cannot lie in the interior of the convex hull any subset of sensors in $\mathcal{C}($C2$)$, and, thus, cannot triangulate themselves, which contradicts Lemma~\ref{lemma:triangulation1} and assumption~{\bf (B3)}. In order to triangulate the boundary sensors in $\mathcal{C}($C2$)$, the boundary sensors in~C2 must be able to reach the transient states and/or the absorbing states, that is to say that the  boundary sensors in $\mathcal{C}($C2$)$ have to reach the sensors in~C1 to be able to triangulate themselves. Hence, the Markov chain is absorbing.
\end{proof}

Consider the partitioning of the iteration matrix, $\mathbf{\Upsilon}$, in~\eqref{Up_part}. With the Markov chain interpretation, the $M\times (m+1)$ block $\mathbf{B}=\{b_{lj}\}$ is a transition probability matrix for the transient states to reach the absorbing states in one-step, and the block $M\times M$ $\mathbf{P}=\{p_{lj}\}$ is a transition probability matrix for the transient states. With~\eqref{Up_part}, $\mathbf{\Upsilon}^{t+1}$ can be written as
\begin{equation}
\label{Upst1}
\mathbf{\Upsilon}^{t+1}=\left[\begin{array}{cc}
\mathbf{I}_{m+1}&\mathbf{0}\\
\displaystyle\sum_{k=0}^{t}\mathbf{P}^{k}\mathbf{B}&\mathbf{P}^{t+1}
\end{array} \right],
\end{equation}
and, as $t$ goes to infinity, we have
\begin{equation}
\label{Upst}
\lim_{t\rightarrow\infty}\mathbf{\Upsilon}^{t+1}=\left[\begin{array}{cc}
\mathbf{I}_{m+1}&\mathbf{0}\\
\left(\mathbf{I}_{M}-\mathbf{P}\right)^{-1}\mathbf{B}&\mathbf{0}
\end{array} \right],
\end{equation}
by Lemmas~\ref{lem2} and~\ref{lem1}, in appendix~\ref{IR}. Lemmas~\ref{lem2} and~\ref{lem1} use the fact that if $\mathbf{P}$ is the matrix associated to the transient states of an absorbing Markov chain, then $\rho(\mathbf{P})<1$, where $\rho(\cdot)$ is the spectral radius of a matrix. With~\eqref{Upst}, DILOC~\eqref{diloc} converges to
\begin{equation}
\label{loc_conv}
\lim_{t\rightarrow\infty}\mathbf{C}(t+1) = \left[\begin{array}{cc}
\mathbf{I}_{m+1}&\mathbf{0}\\
\left(\mathbf{I}_{M}-\mathbf{P}\right)^{-1}\mathbf{B}&\mathbf{0}
\end{array} \right]\mathbf{C}(0).
\end{equation}
In~\eqref{loc_conv}, the coordinates of the~$M$ sensors in~$\Omega$ (last $M$ rows of $\mathbf{C}(t+1)$) are written as a function of the~$m+1$ anchors in~$\kappa$ whose coordinates are exactly known. The limiting values of the states of the~$M$ sensors in~$\Omega$ are written in terms of the coordinates of the~$m+1$ anchors in~$\kappa$ weighted by $(\mathbf{I}_{M}-\mathbf{P})^{-1}\mathbf{B}$. To show that the limiting values are indeed the exact solution, we give the following Lemma.
\begin{lem}\label{lemD}
Let $\mathbf{c}_l^\ast$ be the exact coordinates of a node, $l\in\Theta$. Let the $M\times (m+1)$ matrix, $\mathbf{D}=\{d_{lj}\},l\in\Omega,j\in\kappa$, be the matrix of the barycentric coordinates of the $M$ sensors (in $\Omega$) in terms of the $m+1$ anchors in $\kappa$, relating the coordinates of the sensors to the coordinates of the anchors by \begin{equation}\label{one}\mathbf{c}_l^\ast=\sum_{j\in\kappa} d_{lj} \mathbf{c}_j^\ast,\qquad\qquad l\in\Omega.\end{equation} Then, we have \begin{eqnarray}\label{Dsol}\mathbf{D}&=&\left(\mathbf{I}_{M}-\mathbf{P}\right)^{-1}\mathbf{B}.\end{eqnarray}
\end{lem}
\begin{proof} Clearly $(\mathbf{I}_{M}-\mathbf{P})$ is invertible, since, by~\eqref{lemmm} in Appendix~\ref{IR}, $\rho(\mathbf{P})<1$; this follows from the fact that the eigenvalues of the matrix $\mathbf{I}_{M}-\mathbf{P}$ are $1 - \lambda_j$, where~$\lambda_j$ is the~$j$th eigenvalue of the matrix~$\mathbf{P}$ and $|\lambda_j|<1,~\forall j=1,\ldots,M$. It suffices to show that,
\begin{eqnarray}
\label{conv_eq}
\mathbf{D}&=&\mathbf{B+PD},
\end{eqnarray}
since~\eqref{Dsol} follows from~\eqref{conv_eq}. In~\eqref{conv_eq}, $\mathbf{D}$ and~$\mathbf{B}$ are both $M\times (m+1)$ matrices, whereas~$\mathbf{P}$ is an $M\times M$ matrix whose non-zero elements are the barycentric coordinates for the sensors in~$\Omega$.  Hence, for the~$lj$-th element in~\eqref{conv_eq}, we need to show that
\begin{equation}
\label{dlj}
d_{lj} =  b_{lj} + \sum_{k\in\Omega} p_{lk} d_{kj}.
\end{equation}
For an arbitrary sensor, $l\in\Omega$, its triangulation set, $\Theta_l$, may contain nodes from both~$\kappa$ and~$\Omega$. We denote $\kappa_{\Theta_l}$ as the elements of $\Theta_l$ that are anchors, and $\Omega_{\Theta_l}$ as the elements of $\Theta_l$ that are non-anchor sensors. The exact coordinates, $\mathbf{c}_l^{\ast}$, of the sensor, $l$, can be expressed as a convex combination of the coordinates of its neighbors in its triangulation set, $k\in\Theta_l$, using the barycentric coordinates, $a_{lk}$, i.e.,
\begin{eqnarray}
\label{dij_case1}
\mathbf{c}_l^{\ast} &=& \sum_{k\in\Theta_l}a_{lk}\mathbf{c}_k^{\ast},\nonumber\\&=& \sum_{j\in\kappa_{\Theta_l}}a_{lj}\mathbf{c}_j^{\ast}
+\sum_{k\in\Omega_{\Theta_l}}a_{lk}\mathbf{c}_k^{\ast},\nonumber\\
&=& \sum_{j\in\kappa}b_{lj}\mathbf{c}_j^{\ast}
+\sum_{k\in\Omega}p_{lk}\mathbf{c}_k^{\ast},
\end{eqnarray}
since the scalars, $a_{lj}$, are given by
 \begin{eqnarray}
 a_{lj}=\left\{\begin{array}{cc}b_{lj},&\qquad\mbox{if } j\in\kappa_{\Theta_l},\\p_{lj},&\qquad\mbox{if }j\in\Omega_{\Theta_l},\\0,&\qquad\mbox{if }j\notin\Theta_l.\end{array}\right.
 \end{eqnarray}
 Equation~\eqref{dij_case1} becomes, after writing each $k\in\Omega$ in terms of the~$m+1$ anchors in~$\kappa$,
 \begin{eqnarray}
 \label{lem1_proof}
 \mathbf{c}_l^{\ast} &=& \sum_{j\in\kappa} b_{lj}\mathbf{c}_j^{\ast}+\sum_{k\in\Omega}p_{lk}\sum_{j\in\kappa}d_{kj}\mathbf{c}_j^{\ast},\nonumber\\&=& \sum_{j\in\kappa} b_{lj}\mathbf{c}_j^{\ast}+\sum_{j\in\kappa}\sum_{k\in\Omega}p_{lk}d_{kj}\mathbf{c}_j^{\ast},\nonumber\\&=& \sum_{j\in\kappa}\left( b_{lj}+\sum_{k\in\Omega}p_{lk}d_{kj}\right)\mathbf{c}_j^{\ast}.
 \end{eqnarray}
 This is a representation of the coordinates of sensor, $l$, in terms of the coordinates of the anchors, $j\in\kappa$. Since for each $j\in\kappa$, the value inside the parentheses is non-negative with their sum over $j\in\kappa$ being 1 and the fact that the barycentric representation is unique, we must have
 \begin{equation}
 d_{lj}=b_{lj}+\sum_{k\in\Omega}p_{lk}d_{kj},
 \end{equation}
 which, comparing to~\eqref{one}, completes the proof.
\end{proof}

We now recapitulate these results in the following theorem.
\begin{theorem}[DILOC convergence]
\label{thm_main}
DILOC~\eqref{diloc} converges to the \emph{exact} coordinates, $\mathbf{c}_l^\ast$, of the~$M$ sensors (with unknown locations) in~$\Omega$, i.e.,
\begin{equation}
\lim_{t\rightarrow\infty}\mathbf{c}_l(t+1)=
\mathbf{c}_l^{\ast},\qquad\forall~l\in\Omega.
\end{equation}
\end{theorem}
\begin{proof}
The proof is a consequence of Lemmas~\ref{mp_lem} and~\ref{lemD}.
\end{proof}
%

\textbf{Convergence rate.}
The convergence rate of the localization algorithm depends on the spectral radius of the matrix $\mathbf{P}$, which by~\eqref{lemmm} in Appendix~\ref{IR} is strictly less than one. This is a consequence of the fact that~$\mathbf{P}$ is a sub-stochastic matrix. The convergence is slow if the spectral radius, $\rho(\mathbf{P})$, is close to $1$. This can happen if the matrix $\mathbf{B}$ is close to a zero matrix, $\mathbf{0}$.  This is the case if and only if the sensors cluster in a region of very small area inside the convex hull of the anchors, and the anchors themselves are very far apart. In fact, it can be seen that in this case the barycentric  coordinates for the sensors with $\kappa_{\Theta_l}\neq\varnothing$ (see Lemma~\ref{lemD} for this notation) corresponding to the elements in $\kappa_{\Theta_l}$ are close to zero. Since in practical wireless sensor applications the sensors are assumed to be deployed in a geometric or a Poisson fashion (see details in Section~\ref{rand_dep}), the probability of this to happen is arbitrarily close to~$0$.

\section{DILOC with Relaxation}
\label{rel}
In this Section, we modify DILOC to speed its convergence rate and to obtain a form that is more suitable to study distributed localization in random environments. We observe that in DILOC~\eqref{diloc}, at time $t+1$,
the expression for $\mathbf{c}_l(t+1),~l\in\Omega$, does not
involve its own coordinates, $\mathbf{c}_l(t)$, at time $t$. We introduce a relaxation parameter, $\alpha\in(0,1]$, in the iterations, such that, the
expression of $\mathbf{c}_l(t+1)$ is a convex combination of
$\mathbf{c}_l(t)$ and~\eqref{diloc}. We refer to this version as the DILOC \emph{with relaxation}, DILOC-REL. It is given by
\begin{eqnarray}
\label{diloc_rel}
\mathbf{c}_l(t+1) &=& \left\{\begin{array}{cc}(1-\alpha)\mathbf{c}_l(t)+\alpha \mathbf{c}_l(t)=\mathbf{c}_l(t),&l\in\kappa,\\
(1-\alpha)\mathbf{c}_l(t)
+\alpha\sum_{k\in\Theta_l}a_{lk}\mathbf{c}_k(t),&l\in\Omega.
\end{array}\right.
\end{eqnarray}
DILOC is the special case of DILOC-REL with $\alpha = 1$. The matrix representation of DILOC-REL is
\begin{equation}
\label{loc_OR}
\mathbf{C}(t+1) = \mathbf{HC}(t),
\end{equation}
where $\mathbf{H}=\left(1-\alpha\right)\mathbf{I}_{N}+\alpha\mathbf{\Upsilon}$ and $\mathbf{I}_{N}$ is the $N\times N$ identity matrix.
It is straightforward to show that the iteration matrix, $\mathbf{H}$, corresponds to a transition probability matrix of an absorbing Markov chain, where the anchors are the absorbing states and the sensors are the transient states. Let $\mathbf{J} = \left(1-\alpha\right)\mathbf{I}_{M}+\alpha\mathbf{P}$; partitioning $\mathbf{H}$ as
\begin{equation}
\label{Up_part_or}\mathbf{H}=\left[\begin{array}{cc}
\mathbf{I}_{m+1}&\mathbf{0}\\
\alpha\mathbf{B}&\mathbf{J}\end{array} \right].
\end{equation}
We note the following
\begin{equation}
\label{Upst_OR}
\mathbf{H}^{t+1}=\left[\begin{array}{cc}
\mathbf{I}_{m+1}&\mathbf{0}\\
\displaystyle\sum_{k=0}^{t}\mathbf{J}^{k}\mathbf{\alpha B}&\mathbf{J}^{t+1}
\end{array} \right],
\end{equation}
and, as $t\rightarrow \infty$,
\begin{equation}
\label{ORlim}
\lim_{t\rightarrow\infty}\mathbf{H}^{t+1}=\left[\begin{array}{cc}
\mathbf{I}_{m+1}&\mathbf{0}\\
\left(\mathbf{I}_{M}-\mathbf{J}\right)^{-1}\mathbf{\alpha B}&\mathbf{0}
\end{array} \right],
\end{equation}
from Lemmas~\ref{lem2} and~\ref{lem1}. Lemmas~\ref{lem2} and~\ref{lem1} apply to $\mathbf{H}$, since $\mathbf{H}$ is non-negative and $\rho(\mathbf{J})<1$. To show $\rho(\mathbf{J})<1$, we recall that $\rho(\mathbf{P})<1$ and the eigenvalues of $\mathbf{J}$ are $(1-\alpha)+\alpha\lambda_j$, where $\lambda_j$ are the eigenvalues of $\mathbf{P}$. Therefore, we have
\begin{eqnarray}
\rho(\mathbf{J}) &=& \max_j|(1-\alpha)+\alpha\lambda_j|< 1.
\end{eqnarray}
The following Theorem establishes convergence of DILOC-REL.
 \begin{theorem}
 DILOC-REL~\eqref{diloc_rel} converges to the \emph{exact} coordinates, $\mathbf{c}_l^\ast$, of the~$M$ sensors (with unknown locations) in~$\Omega$, i.e.,
 \begin{equation}
 \lim_{t\rightarrow\infty}\mathbf{c}_l(t+1)=
 \mathbf{c}_l^{\ast},\qquad\forall~l\in\Omega.
 \end{equation}
 \end{theorem}
\begin{proof}
It suffices to show that
\begin{equation}
\label{eq_rel}
\left(\mathbf{I}_{M}-\mathbf{J}\right)^{-1}\mathbf{\alpha B}=\left(\mathbf{I}_{M}-\mathbf{P}\right)^{-1}\mathbf{B}.
\end{equation}
 To this end, we note that
 \begin{eqnarray}
 \left(\mathbf{I}_{M}-\mathbf{J}\right)^{-1}\mathbf{\alpha B}&=&\left(\mathbf{I}_{M}-\left(\left(1-\alpha\right)\mathbf{I}_{M}
 +\alpha\mathbf{P}\right)\right)^{-1}\mathbf{\alpha B},
 \end{eqnarray}
 which reduces to~\eqref{eq_rel} after basic algebraic manipulations. The convergence of DILOC-REL  thus follows from Lemma~\ref{lemD}.
 \end{proof}
As mentioned, the advantage of DILOC-REL is twofold: since $\rho(\mathbf{J})$ is a function of $\alpha$, we may optimize the convergence rate over $\alpha$; and DILOC-REL forms the basis for the distributed localization algorithm in random environments~(DLRE) that we discuss in  Sections~\ref{AlgAss} and~\ref{asconv}.

\section{Distributed Localization in Random Environments: Assumptions and Algorithm} \label{AlgAss}
 This and the next Section study distributed iterative localization in more realistic practical scenarios, when the inter-sensor distances are known up to errors, the communication links between sensors may fail and, when alive, the communication among sensors is corrupted by noise.
 We write the update equations for DILOC-REL,~(\ref{loc_OR}), in terms of the columns,~$\mathbf{c}^{j}(t)$,~$1\leq j\leq m$, of the coordinate matrix,~$\mathbf{C}(t)$. Column~$j$ corresponds to the vector of  the~$j$-th estimate coordinates of all the~$N$ sensor locations\footnote{In the sequel, we omit the subscripts from the identity matrix,~$\mathbf{I}$, and its dimensions will be clear from the context.}. The updates are
\begin{equation}
\label{algass:1} \mathbf{c}^{j}(t+1)=\left[(1-\alpha
)\mathbf{I}+\alpha\mathbf{\Upsilon}\right]\mathbf{c}^{j}(t),~1\leq j\leq m.
\end{equation}
We partition~$\mathbf{c}^{j}(t)$ as
\begin{equation}
\label{algass:2} \mathbf{c}^{j}(t) = \left[ \begin{array}{ll}
                    \mathbf{u}^{j} \\
                    \mathbf{x}^{j}(t)
                   \end{array}
          \right],
\end{equation}
where,~$\mathbf{u}^{j}\in\mathbb{R}^{(m+1)\times 1}$ corresponds to
the~$j$-th coordinates of the anchors, which are know (hence, we omit the time index, as they
are not updated), and~$\mathbf{x}^{j}(t)\in\mathbb{R}^{M\times 1}$
corresponds to the estimates of the~$j$-th coordinates of the non-anchor sensors, hence not known. Since, update is performed only on the~$\mathbf{x}^{j}(t)$,~(\ref{algass:1}) is equivalent to the following recursion:
\begin{equation}
\label{algass:3}
\mathbf{x}^{j}(t+1)=\left[(1-\alpha)\mathbf{I}+\alpha
\mathbf{P}\right]\mathbf{x}^{j}(t)+\alpha \mathbf{B}\mathbf{u}^{j}.
\end{equation}
Thus, to implement the sequence of iterations in~(\ref{algass:3}) perfectly, the~$l$-th sensor at iteration~$t$ needs the corresponding rows of the matrices~$\mathbf{P}$ and~$\mathbf{B}$, and, in addition, the current estimates,
~$c_{n}^{j}(t), n\in\Theta_{l}$ ($j$-th component of the~$n$-th sensor coordinates), of its neighbors' positions.  In practice, there are several limitations:
 \begin{inparaenum}[(i)]
 \item The computation of the matrices~$\mathbf{P}$ and~$\mathbf{B}$ requires inter-sensor distance computations, which are not perfect in a random environment; \item the communication channels, or links, between neighboring channels may fail at random times; and \item because of imperfect communication, each sensor receives only noisy versions of its neighbors current state. \end{inparaenum} Hence, in a random environment, we need to modify the iteration sequence in~(\ref{algass:3}) to account for the partial imperfect information received by a sensor at each iteration. We start by stating formally our modeling assumptions.
\begin{itemize}[\setlabelwidth{(1)}]
\item{\textbf{(C1)~Randomness in system matrices}}. At each iteration, each sensor needs the corresponding row of the system matrices~$\mathbf{B}$ and~$\mathbf{P}$, which in turn, depend on the inter-sensor distance measurements, which can be, possibly, random. Since a single measurement of the inter-sensor distances may lead to a large random noise, we assume the sensors estimate the required distances at each iteration of the algorithm (note that this leads to an implicit averaging of the unbiased noisy effects, as will be demonstrated later.) In other words, at each
iteration, the~$l$-th sensor can only get estimates,~$\widehat{\mathbf{B}}_{l}(t)$ and~$\widehat{\mathbf{P}}_{l}(t)$, of the corresponding rows of the~$\mathbf{B}$ and~$\mathbf{P}$ matrices, respectively. In the generic imperfect communication case, we have
\begin{equation}
\label{algass:4}
\widehat{\mathbf{B}}(t)=\mathbf{B}+\mathbf{S}_{\mathbf{B}}
+\widetilde{\mathbf{S}}_{\mathbf{B}}(t)
\end{equation}
where~$\left\{\widetilde{\mathbf{S}}_{\mathbf{B}}(t)\right\}_{t\geq 0}$ is an independent sequence of random matrices with,
\begin{equation}
\label{algass:5} \mathbb{E}\left[\widetilde{\mathbf{S}}_{\mathbf{B}}(t)\right]=0,\:\forall t, \:\:\sup_{t\geq 0} \mathbb{E}\left[\left\|\widetilde{\mathbf{S}}_{\mathbf{B}}(t)\right\|^{2}\right]
=k_{\mathbf{B}}<\infty.
\end{equation}
Here,~$\mathbf{S}_{\mathbf{B}}$ is the mean measurement error. Similarly, for~$\mathbf{P}$, we have
\begin{equation}
\label{algass:6} \widehat{\mathbf{P}}(t)=\mathbf{P}+\mathbf{S}_{\mathbf{P}}
+\widetilde{\mathbf{S}}_{\mathbf{P}}(t),
\end{equation}
where~$\left\{\widetilde{\mathbf{S}}_{\mathbf{P}}(t)\right\}_{t\geq 0}$ is an independent sequence of random matrices with,
\begin{equation}
\label{algass:7}
\mathbb{E}\left[\widetilde{\mathbf{S}}_{\mathbf{P}}(t)\right]=0,\:\forall t, \:\:\sup_{t\geq 0} \mathbb{E}\left[\left\|\widetilde{\mathbf{S}}_{\mathbf{P}}(t)\right\|^{2}\right]
=k_{\mathbf{P}}<\infty.
\end{equation}
Likewise,~$\mathbf{S}_{\mathbf{P}}$ is the mean measurement error. Note that this way of writing~$\widetilde{B}(t),\widetilde{P}(t)$
does not require the noise model to be additive. It only says that
any random object may be written as the sum of a deterministic
mean part and the corresponding zero mean random part. The moment
assumptions in eqns.~(\ref{algass:5},\ref{algass:7}) are very weak
and, in particular, are satisfied if the sequences~$\left\{\widehat{\mathbf{B}}(t)\right\}_{t\geq 0}$ and
~$\left\{\widehat{\mathbf{P}}(t)\right\}_{t\geq 0}$ are i.i.d.~with finite variance.
\item{\textbf{(C2)~Random Link Failure}}: We assume that the inter-sensor communication links fail randomly. This happens, for example, in wireless sensor network applications, where occasionally data packets are dropped. To this end, if the sensors~$l$ and~$n$ share a communication link (or,~$n\in\Theta_{l}$), we assume that the link fails with some probability~$1-q_{ln}$ at each iteration, where~$0<q_{ln}\leq 1$. We associate with each such potential network link, a binary random variable,~$e_{ln}(t)$, where~$e_{ln}(t)=1$ indicates that the corresponding network link is active at time~$t$, whereas~$e_{ln}(t)=0$ indicates a link failure. Note that~$\mathbb{E}\left[e_{ln}\right]=q_{ln}$.
\item{\textbf{(C3)~Additive Channel Noise}}: Let~$\left\{v_{ln}^{j}(t)\right\}_{l,n,j,t}$ be a family of independent zero mean random variables such that
\begin{equation}
\label{algass:8}
\sup_{l,n,j,t}\mathbb{E}\left[v_{ln}^{j}(t)\right]^{2}=k_{v}<\infty.
\end{equation}
We assume that, at the~$t$-th iteration, if the network link
~$(l,n)$ is active, sensor~$l$ receives only a corrupt version,
~$y_{ln}^{j}(t)$, of sensor~$n$'s state,~$c_{n}^{j}(t)$, given by
\begin{equation}
\label{algass:9} y_{ln}^{j}(t)=c_{n}^{j}(t)+v_{ln}^{j}(t).
\end{equation}
This models the channel noise. The moment assumption in eqn.~(\ref{algass:8}) is very weak and holds, in particular, if the channel noise is i.i.d.

\item{\textbf{(C4)~Independence}}: We assume that the sequences,
~$\left\{\widetilde{\mathbf{S}}_{\mathbf{B}}(t),
\widetilde{\mathbf{S}}_{\mathbf{P}}(t)\right\}_{t\geq 0}$,
~$\left\{e_{ln}(t)\right\}_{l,n,t}$, and~$\left\{v_{ln}^{j}(t)\right\}_{l,n,j,t}$ are mutually independent. These assumptions do not put restrictions on the distributional form of the random errors, only that they obey some weak moment conditions.
\end{itemize}
Clearly, under the random environment model (as detailed in Assumptions~\textbf{(C1)-(C4)}, the algorithm in~(\ref{algass:3}) is not appropriate to update the sensors states.  We now consider the following state update recursion for the random environment case.

\textbf{Distributed Localization in Random Environment Algorithm (DLRE):}
\begin{eqnarray}
\label{algass:10}
x_{l}^{j}(t+1)&=&\left(1-\alpha\left(t\right)\right)x_{l}^{j}(t)+
\alpha(t)\left[\sum_{n\in\kappa\cap\Theta_{l}}
\frac{e_{ln}(t)\widehat{\mathbf{B}}_{ln}(t)}{q_{ln}}\left(u_{n}^{j} +v_{ln}^{j}(t)\right)\right]\\
\nonumber
&+&
\alpha(t)\left[\sum_{n\in\Omega\cap\Theta_{l}}
\frac{e_{ln}(t)\widehat{\mathbf{P}}_{ln}(t)}{q_{ln}}\left(x_{n}^{j}(t)
+v_{ln}^{j}(t)\right)\right],\:\:l\in\Omega,\,\,1\leq j\leq m
\end{eqnarray}
In contrast with DILOC-REL, in~(\ref{algass:10}), the gain~$\alpha(t)$ is now time varying. It will become clear why when we study convergence of this algorithm. To write DLRE in a compact form, we introduce notation. Define the random matrices,~$\widetilde{\mathbf{B}}(t)\in\mathbb{R}^{M\times (m+1)}$ and
~$\widetilde{\mathbf{P}}(t)\in\mathbb{R}^{M\times (m+1)}$, as the matrices with~$ln$ entries given by
\begin{equation}
\label{algass:11}
\widetilde{\mathbf{B}}_{ln}(t) =
\widehat{\mathbf{B}}_{ln}(t)
\left(\frac{e_{ln}(t)}{q_{ln}}-1\right),\:\:\widetilde{\mathbf{P}}_{ln}(t)
=
\widehat{\mathbf{P}}_{ln}(t)\left(\frac{e_{ln}(t)}{q_{ln}}-1\right).
\end{equation}
Clearly, by~\textbf{(C2),(C4)}, the matrices
~$\widetilde{\mathbf{B}}(t)\in\mathbb{R}^{M\times (m+1)}$ and
~$\widetilde{\mathbf{P}}(t)\in\mathbb{R}^{M\times (m+1)}$ are zero mean. Note that~$\mathbb{E}\left[e_{ln}\right]=q_{ln}$. Also,
by the bounded moment assumptions in~\textbf{(C1)}, we have
\begin{equation}
\label{algass:12} \sup_{t\geq 0} \mathbb{E}\left[
\left\|\widetilde{\mathbf{B}}(t)\right\|^{2}\right]=
\widetilde{k}_{\mathbf{B}}<\infty,\:\:\sup_{t\geq 0} \mathbb{E}\left[\left\|\widetilde{\mathbf{P}}(t)\right\|^{2}\right]
=\widetilde{k}_{\mathbf{P}}<\infty.
\end{equation}
Hence, the iterations in~(\ref{algass:10}) can be written in vector form as
\begin{equation}
\label{algass:13}
\mathbf{x}^{j}(t+1)=\left(1-\alpha(t)\right)
\mathbf{x}^{j}(t)+\alpha(t)\left[\left(\widehat{\mathbf{P}}(t)+
\widetilde{\mathbf{P}}(t)\right)\mathbf{x}^{j}(t)
+\left(\widehat{\mathbf{B}}(t)+
\widetilde{\mathbf{B}}(t)\right)\mathbf{u}^{j}+\mathbf{\eta}^{j}(t)\right],
\end{equation}
where, the~$l$th element of the vector,~$\mathbf{\eta}^{j}(t)$, is given by
\begin{equation}
\label{algass:14} \mathbf{\eta}_{l}^{j}(t)=\sum_{n\neq l} \left(\widehat{\mathbf{P}}_{ln}(t)
+\widetilde{\mathbf{P}}_{ln}(t)\right)v_{ln}^{j}(t)
+\sum_{n\neq l}
\left(\widehat{\mathbf{B}}_{ln}(t)
+\widetilde{\mathbf{B}}_{ln}(t)\right)v_{ln}^{j}(t).
\end{equation}
By~\textbf{(C1)-(C4)}, the sequence,~$\{\mathbf{\eta}^{j}(t)\}_{t\geq
0}$, is zero mean, independent, with
\begin{equation}
\label{algass:15}
\sup_{t}\mathbb{E}\left[\left\|\mathbf{\eta}^{j}(t)\right\|^{2}\right]
=k_{\eta}<\infty.
\end{equation}
From~\textbf{(C1)}, the iteration sequence in~(\ref{algass:13}) can be written as
\begin{eqnarray}
\label{algass:16}
\mathbf{x}^{j}(t+1)&=&\mathbf{x}^{j}(t)-
\alpha(t)\left[\left(\mathbf{I}-\mathbf{P}-
\mathbf{S}_{\mathbf{P}}\right)\mathbf{x}^{j}(t)-\left(\mathbf{B}+ \mathbf{S}_{\mathbf{B}}\right)\mathbf{u}^{j}
-\left(\widetilde{\mathbf{S}}_{\mathbf{P}}(t)
+\widetilde{\mathbf{P}}(t)\right)\mathbf{x}^{j}(t)\right.\\\nonumber
& &
\left.-\left(\widetilde{\mathbf{S}}_{\mathbf{B}}(t)
+\widetilde{\mathbf{B}}(t)\right)\mathbf{u}^{j}
-\mathbf{\eta}^{j}(t)\right].
\end{eqnarray}
We now make two additional design assumptions.
\begin{itemize}[\setlabelwidth{(1)}]
\item{\textbf{(D1)~Persistence Condition}}: The weight sequence satisfies
\begin{equation}
\label{alphacond} \alpha (t)>0,\:\:\sum_{t\geq 0}\alpha
(t)=\infty,\:\:\sum_{t\geq 0}\alpha^{2}(t)<\infty.
\end{equation}
This condition, commonly assumed in the adaptive control and adaptive signal processing literature, assumes that the weights decay to zero, but not too fast.
\item{\textbf{(D2)~Low Error Bias}}: We assume that
\begin{equation}
\label{algass:17}
\rho\left(\mathbf{P}+\mathbf{S}_{\mathbf{P}}\right)<1.
\end{equation}
Clearly, we have~$\rho(\mathbf{P})<1$. Thus, if we assume that the non-zero bias,~$\mathbf{S}_{\mathbf{P}}$, in the system matrix (resulting from incorrect distant computation) is small,~(\ref{algass:17}) is justified. We note that this condition ensures that the matrix
~$\left(\mathbf{I}-\mathbf{P}-\mathbf{S}_{\mathbf{P}}\right)$ is invertible.
\end{itemize}
In the following sections, we prove that the DLRE algorithm, under the assumptions~\textbf{(C1)-(C4), (D1)-(D2)}, leads to a.s.~convergence of the state vector sequence,~$\left\{\mathbf{x}^{j}(t)\right\}_{t\geq 0}$, to a deterministic vector for each~$j$, which may be different from the exact sensor locations, because of the random errors in the iterations. We characterize this error and show that it depends on the non-zero biases,~$\mathbf{S}_{\mathbf{B}}$, and~$\mathbf{S}_{\mathbf{P}}$ in the system matrix computations, and vanishes as
~$\left\|\mathbf{S}_{\mathbf{B}}\right\|\rightarrow 0$ and~$\left\|\mathbf{S}_{\mathbf{P}}\right\|\rightarrow 0$.

\section{DLRE: A.S.~Convergence}\label{as_conv}
\label{asconv}
\label{MarkProc}
We show the almost sure convergence of DLRE  under the random environment presented in Section~\ref{AlgAss}.
\begin{theorem}[DLRE a.s.~convergence]
\label{conv} Let~$\{\mathbf{x}^{j}(t)\}_{t\geq 0},~1\leq j\leq m$,
be the state sequence generated by the iterations, given by
~(\ref{algass:16}), under the assumptions~\textbf{(C1)-(C4),
(D1)-(D2)}. Then,
\begin{equation}
\label{th:conv1}
\mathbb{\mathbf{P}}\left[\lim_{t\rightarrow\infty}\mathbf{x}^{j}(t)=\left(\mathbf{I}-\mathbf{P}-\mathbf{S}_{\mathbf{P}}\right)^{-1} \left(\mathbf{B}+\mathbf{S}_{\mathbf{B}}\right)\mathbf{u}^{j},~\forall j\right]=1.
\end{equation}
\end{theorem}
The convergence analysis of the DLRE algorithm is based on the
sample path properties of controlled Markov processes, which has
also been used recently to prove convergence properties of
distributed iterative stochastic algorithms in sensor networks,
e.g., \cite{karmoura-randomtopologynoise,karmoura-quantized}. The
proof relies on the following result from~\cite{Nevelson}, which
we state here as a theorem.
\begin{theorem}
\label{RM} Consider the following recursive procedure:
\begin{equation}
\label{RM:1}
\mathbf{x}(t+1)=\mathbf{x}(t)+\alpha(t)\left[\mathbf{R}(\mathbf{x}(t))+\mathbf{\Gamma}
(t+1,\mathbf{x}(t),\omega)\right],
\end{equation}
where,~$\mathbf{x},\mathbf{R},\mathbf{\Gamma}$ are vectors
in~$\mathbb{{R}}^{M\times 1}$. There is an underlying common probability
space~$(\mathbf{\Xi},\mathcal{F},\mathcal{\mathbf{P}})$, and
~$\omega$ is the canonical element of the probability space,
~$\mathbf{\Xi}$. Assume that the following conditions are
satisfied\footnote{In the sequel,~$\mathcal{B}^{M}$ denotes the
Borel sigma algebra in~$\mathbb{R}^{M\times 1}$. The space of
twice continuously differentiable functions is denoted by
~$\mathcal{C}_{2}$, while~$V_{\mathbf{x}}(\mathbf{x})$ denotes the
gradient~$\frac{\partial V(\mathbf{x})}{\partial\mathbf{x}}$.}.
\begin{itemize}[\setlabelwidth{1)}]
\item{\textbf{1)}}: The vector function~$\mathbf{R}(\mathbf{x})$ is Borel measurable and~$\mathbf{\Gamma}(t,\mathbf{x},\omega)$ is
~$\mathcal{B}^{M}\otimes\mathcal{F}$ measurable for every~$t$.
\item{\textbf{2)}}: There exists a filtration
~$\{\mathcal{F}_{t}\}_{t\geq 0}$ of~$\mathcal{F}$, such that the
family of random vectors~$\mathbf{\Gamma}(t,\mathbf{x},\omega)$ is
~$\mathcal{F}_{t}$ measurable, zero-mean and independent of
~$\mathcal{F}_{t-1}$.
\item{\textbf{3)}}: There exists a function
~$V(\mathbf{x})\in\mathbb{C}^{2}$ with bounded second order partial
derivatives satisfying:
\begin{eqnarray}
\label{Vcond:1}
V(\mathbf{x}_{0})=0,\:\:V(\mathbf{x})&>&0,\:\mathbf{x}\neq\mathbf{x}_{0},\\
\label{Vcond:2}
\sup_{\|\mathbf{x}-\mathbf{x}_{0}\|>\epsilon}
\left<\mathbf{R}(\mathbf{x}),V_{\mathbf{x}}(\mathbf{x})\right>
&<&0,\:\:\forall\epsilon>0.
\end{eqnarray}
\item{\textbf{4)}}: There exist constants~$k_{1},k_{2}>0$, such
that,
\begin{equation}
\label{Lcond}
\|\mathbf{R}(\mathbf{x})\|^{2}+\mathbb{E}[\|\mathbf{\Gamma}(t,\mathbf{x},\omega)\|^{2}]\leq
k_{1}(1+V(\mathbf{x}))-k_{2}\left<\mathbf{R}(\mathbf{x}),V_{\mathbf{x}}(\mathbf{x})\right>
\end{equation}
\item{\textbf{5)}}: The weight sequence~$\{\alpha(t)\}_{t\geq 0}$
satisfies the persistence condition~\textbf{(D1)} given by
~(\ref{alphacond}).
%
\end{itemize}
Then the Markov process,~$\{\mathbf{x}(t)\}_{t\geq 0}$, converges
a.s. to~$\mathbf{x}_{0}$.
\end{theorem}
\begin{proof}
The proof follows from Theorem 4.4.4 in~\cite{Nevelson} and is
omitted due to space constraints.
\end{proof}

We now return to the proof of Theorem~\ref{conv}.

\begin{proof}[Proof of Theorem~\ref{conv}]
We will show that, under the assumptions, the algorithm in
~(\ref{algass:16}) falls under the purview of Theorem~\ref{RM}. To
this end, consider the filtration,~$\{\mathcal{F}_{t}\}_{t\geq
0}$, where
\begin{equation}
\label{th:conv2}
\mathcal{F}_{t}=\sigma\left(\mathbf{x}^{j}(0),\widetilde{\mathbf{S}}_{\mathbf{P}}(s),\widetilde{\mathbf{P}}(s), \widetilde{\mathbf{S}}_{\mathbf{B}}(s),\widetilde{\mathbf{B}}(s),\mathbf{\eta}^{j}(s):~0\leq s<t\right).
\end{equation}
Define the vector~$\mathbf{d}^{\ast}$ as
\begin{equation}
\label{th:conv3}
\mathbf{d}^{\ast}=\left(\mathbf{I}-\mathbf{P}-\mathbf{S}_{\mathbf{P}}\right)^{-1}\left(\mathbf{B}+\mathbf{S}_{\mathbf{B}}\right)\mathbf{u}^{j}.
\end{equation}
Equation~(\ref{algass:16}) can be written as
\begin{eqnarray}
\label{th:conv4} \mathbf{x}^{j}(t+1)=\mathbf{x}^{j}(t)&-&\alpha(t)
\left[\left(\mathbf{I}-\mathbf{P}-\mathbf{S}_{\mathbf{P}}\right)\left(\mathbf{x}^{j}(t)-\mathbf{d}^{\ast}\right)\nonumber
-\left(\widetilde{\mathbf{S}}_{\mathbf{P}}(t)+\widetilde{\mathbf{P}}(t)\right)\mathbf{x}^{j}(t)\right.\\&-&\left.\left(\widetilde{\mathbf{S}}_{\mathbf{B}}(t)+
\widetilde{\mathbf{B}}(t)\right)\mathbf{u}^{j}-\mathbf{\eta}^{j}(t)\right].
\end{eqnarray}
In the notation of Theorem~\ref{RM},~(\ref{th:conv4}) is given by
\begin{equation}
\label{th:conv5}
\mathbf{x}^{j}(t+1)=\mathbf{x}^{j}(t)+\alpha(t)\left[\mathbf{R}(\mathbf{x}^{j}(t))+\mathbf{\Gamma}
(t+1,\mathbf{x}^{j}(t),\omega)\right],
\end{equation}
where
\begin{equation}
\label{th:conv6}
\mathbf{R}(\mathbf{x})=-\left(\mathbf{I}-\mathbf{P}-\mathbf{S}_{\mathbf{P}}\right)\left(\mathbf{x}^{j}(t)-\mathbf{d}^{\ast}\right),
\end{equation}
and
\begin{equation}
\label{th:conv7}
\mathbf{\Gamma}(t+1,\mathbf{x},\omega)=\left[\left(\widetilde{\mathbf{S}}_{\mathbf{P}}(t)+\widetilde{\mathbf{P}}(t)\right)\mathbf{x}^{j}(t)+
\left(\widetilde{\mathbf{S}}_{\mathbf{B}}(t)+\widetilde{\mathbf{B}}(t)\right)\mathbf{u}^{j}+\mathbf{\eta}^{j}(t)\right].
\end{equation}
This definition satisfies assumptions \textbf{1)} and~\textbf{2)}
of Theorem~\ref{RM}.

We now show the existence of a stochastic potential function
~$V(\cdot)$ satisfying the remaining assumptions of
Theorem~\ref{RM}. To this end, define
\begin{equation}
\label{th:conv8}
V(\mathbf{x})=\|\mathbf{x}-\mathbf{d}^{\ast}\|^{2}.
\end{equation}
Clearly,~$V(\mathbf{x})\in\mathbb{C}_{2}$ with bounded second
order partial derivatives, with
\begin{equation}
\label{th:conv9}
V(\mathbf{d}^{\ast})=0,~~V(\mathbf{x})>0,~\mathbf{x}\neq\mathbf{d}^{\ast}.
\end{equation}
Also, we note that, for~$\epsilon>0,$
\begin{eqnarray}
\label{th:conv10}
\sup_{\|\mathbf{x}-\mathbf{d}_{\ast}\|>\epsilon}\left(\mathbf{R}(\mathbf{x}),V_{\mathbf{x}}(\mathbf{x})\right)
& = & \sup_{\|\mathbf{x}-\mathbf{d}_{\ast}\|>\epsilon}-
2\left(\mathbf{x}-\mathbf{d}^{\ast}\right)^{T}\left(\mathbf{I}-\mathbf{P}-\mathbf{S}_{\mathbf{P}}\right)\left(\mathbf{x}-\mathbf{d}^{\ast}\right)\nonumber,
\\ & = &
\sup_{\|\mathbf{x}-\mathbf{d}_{\ast}\|>\epsilon}\left[2\left(\mathbf{x}-\mathbf{d}^{\ast}\right)^{T}\left(\mathbf{P}+\mathbf{S}_{\mathbf{P}}\right)\left(\mathbf{x}-\mathbf{d}^{\ast}\right)-2\|\mathbf{x}-\mathbf{d}^{\ast}\|^{2}\right]\nonumber,
\\ & \leq &
\sup_{\|\mathbf{x}-\mathbf{d}_{\ast}\|>\epsilon}\left[2\left|\left(\mathbf{x}-\mathbf{d}^{\ast}\right)^{T}\left(\mathbf{P}+\mathbf{S}_{\mathbf{P}}\right)\left(\mathbf{x}-\mathbf{d}^{\ast}\right)\right|-2\|\mathbf{x}-\mathbf{d}^{\ast}\|^{2}\right]\nonumber,
\\ & \leq &
\sup_{\|\mathbf{x}-\mathbf{d}_{\ast}\|>\epsilon}\left[2\left\|\mathbf{x}-\mathbf{d}^{\ast}\right\|\rho\left(\mathbf{P}+\mathbf{S}_{\mathbf{P}}\right)\left\|\mathbf{x}-\mathbf{d}^{\ast}\right\|-2\|\mathbf{x}-\mathbf{d}^{\ast}\|^{2}\right]\nonumber,
\\ & = &
\sup_{\|\mathbf{x}-\mathbf{d}_{\ast}\|>\epsilon}-2\left(1-\rho\left(\mathbf{P}+\mathbf{S}_{\mathbf{P}}\right)\right)\|\mathbf{x}-\mathbf{d}^{\ast}\|^{2}\nonumber,
\\ & \leq &
-2\epsilon^{2}\left(1-\rho\left(\mathbf{P}+\mathbf{S}_{\mathbf{P}}\right)\right)\nonumber, \\
& < & 0,
\end{eqnarray}
where, the last step follows from~\textbf{(D2)}. Thus,
assumption~\textbf{3)} in Theorem~\ref{RM} is satisfied.

To verify~\textbf{4)} note that
\begin{eqnarray}
\label{th:conv11} \left\|\mathbf{R}(\mathbf{x})\right\|^{2} & = &
\left(\mathbf{x}-\mathbf{d}^{\ast}\right)^{T}\left(\mathbf{I}
-\mathbf{P}-\mathbf{S}_{\mathbf{P}}\right)^{T}
\left(\mathbf{I}-\mathbf{P}-\mathbf{S}_{\mathbf{P}}\right)
\left(\mathbf{x}-\mathbf{d}^{\ast}\right)\nonumber,
\\ & \leq &
\left\|\left(\mathbf{I}-\mathbf{P}-\mathbf{S}_{\mathbf{P}}\right)^{T}
\left(\mathbf{I}-\mathbf{P}-\mathbf{S}_{\mathbf{P}}\right)\right\|
\left\|\mathbf{x}-\mathbf{d}^{\ast}\right\|^{2}\nonumber,
\\ & = & k_{1}\left\|\mathbf{x}-\mathbf{d}^{\ast}\right\|^{2}\nonumber, \\
& = & k_{1}V(\mathbf{x}),
\end{eqnarray}
where~$k_{1}>0$ is a constant.

Finally, by assumptions~\textbf{(C1)-(C4)}, we have
{\small
\begin{eqnarray}
\label{th:conv12}
\mathbb{E}\left\|\mathbf{\Gamma}(t,\mathbf{x},\omega)\right\|^{2}
& = &
\mathbb{E}\left[\left(\widetilde{\mathbf{S}}_{\mathbf{P}}(t-1)+
\widetilde{\mathbf{P}}(t-1)\right)\mathbf{x}
+\left(\widetilde{\mathbf{S}}_{\mathbf{B}}(t-1)
+\widetilde{\mathbf{B}}(t-1)\right)\mathbf{u}^{j}
+\mathbf{\eta}^{j}(t-1)\right]^{T}
\nonumber
\\
& &
\left[\left(\widetilde{\mathbf{S}}_{\mathbf{P}}(t-1)
+\widetilde{\mathbf{P}}(t-1)\right)\mathbf{x}
+\left(\widetilde{\mathbf{S}}_{\mathbf{B}}(t-1)
+\widetilde{\mathbf{B}}(t-1)\right)\mathbf{u}^{j}
+\mathbf{\eta}^{j}(t-1)\right],
\nonumber
\\
& = &
\mathbf{x}^{T}\mathbb{E}\left[\widetilde{\mathbf{S}}^{T}_{\mathbf{P}}(t-1)
\widetilde{\mathbf{S}}_{\mathbf{P}}(t-1)
+\widetilde{\mathbf{P}}^{T}(t-1)\widetilde{\mathbf{P}}(t-1)\right]\mathbf{x}
\nonumber
 +\mathbf{u}^{jT}\mathbb{E}\left[\widetilde{\mathbf{S}}^{T}_{\mathbf{B}}(t-1)
\widetilde{\mathbf{S}}_{\mathbf{B}}(t-1)\right.
\\
&+&\left.\widetilde{\mathbf{B}}^{T}(t-1)
\widetilde{\mathbf{B}}(t-1)\right]\mathbf{u}^{j}
+\mathbb{E}\left[\left\|\mathbf{\eta}^{j}(t-1)\right\|^{2}\right]
+2\mathbf{x}^{T}\mathbb{E}
\left[\widetilde{\mathbf{S}}^{T}_{\mathbf{P}}(t-1)
\widetilde{\mathbf{S}}_{\mathbf{B}}(t-1)\right]
\mathbf{u}^{j}
\nonumber,
\\
& \leq &
\mathbb{E}\left[\left\|\widetilde{\mathbf{S}}_{\mathbf{P}}(t-1)\right\|^{2}
+\left\|\widetilde{\mathbf{P}}(t-1)\right\|^{2}\right]
\left\|\mathbf{x}\right\|^{2}
+\mathbb{E}\left[\left\|\widetilde{\mathbf{S}}_{\mathbf{B}}(t-1)\right\|^{2}
+\left\|\widetilde{\mathbf{B}}(t-1)\right\|^{2}\right]
\left\|\mathbf{u}^{j}\right\|^{2}
\nonumber
\\
& &
+\mathbb{E}\left[\left\|\mathbf{\eta}^{j}(t-1)\right\|^{2}\right]
+2\mathbb{E}\left[\left\|\widetilde{\mathbf{S}}_{\mathbf{P}}(t-1)
\right\|^{2}\right]^{1/2}
\mathbb{E}\left[\left\|\widetilde{\mathbf{S}}_{\mathbf{B}}(t-1)\right\|^{2}
\right]^{1/2}\left\|\mathbf{x}\right\|\left\|\mathbf{u}^{j}\right\|
\nonumber,
\\
& \leq &
\left(k_{\mathbf{P}}+\widetilde{k}_{\mathbf{P}}\right)
\left\|\mathbf{x}\right\|^{2}
+\left(k_{\mathbf{B}}+\widetilde{k}_{\mathbf{B}}\right)
\left\|\mathbf{u}^{j}\right\|^{2}+k_{\eta}
+k_{\mathbf{P}}k_{\mathbf{B}}\left\|\mathbf{x}\right\|\left\|\mathbf{u}^{j}\right\|.
\end{eqnarray}
}
The cross terms dropped in the second step of
eqn.~(\ref{th:conv12}) have zero mean by the independence
assumption~\textbf{(C4)}. For example, consider the term
~$\mathbb{E}\left[\widetilde{\mathbf{S}}^{T}_{\mathbf{P}}(t-1)
\widetilde{\mathbf{P}}(t-1)\right]$.
It follows from eqns.~(\ref{algass:6},\ref{algass:11}) that the
~$ln$-th entry of the matrix
~$\widetilde{\mathbf{S}}^{T}_{\mathbf{P}}(t-1)\widetilde{\mathbf{P}}(t-1)$
is given by
{\small
\begin{eqnarray}
\label{reply1}
\left[\widetilde{\mathbf{S}}^{T}_{\mathbf{P}}(t-1)
\widetilde{\mathbf{P}}(t-1)\right]_{ln}&
= &
\sum_{r}\left[\widetilde{\mathbf{S}}^{T}_{\mathbf{P}}(t-1)\right]_{lr}
\left[\widetilde{\mathbf{P}}(t-1)\right]_{rn}
\nonumber
\\
\nonumber
 & = &
\sum_{r}\left[\widetilde{\mathbf{S}}_{\mathbf{P}}(t-1)\right]_{rl}
\left[\widehat{\mathbf{P}}(t-1)\right]_{rn}
\left(\frac{e_{rn}(t)}{q_{rn}}-1\right)\nonumber
\\
\nonumber
& = &
\sum_{r}\left[\widetilde{\mathbf{S}}_{\mathbf{P}}(t-1)\right]_{rl}
\left[\mathbf{P}\right]_{rn}
\left(\frac{e_{rn}(t)}{q_{rn}}-1\right)
+\sum_{r}\left[\widetilde{\mathbf{S}}_{\mathbf{P}}(t-1)\right]_{rl}
\left[\mathbf{S}_{\mathbf{P}}\right]_{rn}
\left(\frac{e_{rn}(t)}{q_{rn}}-1\right)
\\&+&\sum_{r}\left[\widetilde{\mathbf{S}}_{\mathbf{P}}(t-1)\right]_{rl}
\left[\widetilde{\mathbf{S}}_{\mathbf{P}}(t-1)\right]_{rn}
\left(\frac{e_{rn}(t)}{q_{rn}}-1\right)
\end{eqnarray}
}
From the independence and zero-mean assumptions, we have the
following,~$\forall r$:
{\scriptsize
\begin{eqnarray}
\label{reply2}
\mathbb{E}\left[\left[\widetilde{\mathbf{S}}_{\mathbf{P}}(t-1)\right]_{rl}
\left[\mathbf{P}\right]_{rn}\left(\frac{e_{rn}(t)}{q_{rn}}-1\right)\right]
& = &
\left[\mathbf{P}\right]_{rn}\mathbb{E}\left[
\left[\widetilde{\mathbf{S}}_{\mathbf{P}}(t-1)\right]_{rl}\right]
\mathbb{E}\left[\left(\frac{e_{rn}(t)}{q_{rn}}-1\right)\right]\nonumber
\\
\nonumber
& = & 0\\
\label{reply3}
\mathbb{E}\left[\left[\widetilde{\mathbf{S}}_{\mathbf{P}}(t-1)\right]_{rl}
\left[\mathbf{S}_{\mathbf{P}}\right]_{rn}
\left(\frac{e_{rn}(t)}{q_{rn}}-1\right)\right]
& = &
\left[\mathbf{S}_{\mathbf{P}}\right]_{rn}
\mathbb{E}\left[\left[\widetilde{\mathbf{S}}_{\mathbf{P}}(t-1)\right]_{rl}
\right]\mathbb{E}\left[\left(\frac{e_{rn}(t)}{q_{rn}}-1\right)\right]
\nonumber
\\
\nonumber
& = & 0\\
\label{reply4}
\mathbb{E}\left[\left[\widetilde{\mathbf{S}}_{\mathbf{P}}(t-1)\right]_{rl}
\left[\widetilde{\mathbf{S}}_{\mathbf{P}}(t-1)
\right]_{rn}\left(\frac{e_{rn}(t)}{q_{rn}}-1\right)\right]
& = &
\mathbb{E}\left[\left[\widetilde{\mathbf{S}}_{\mathbf{P}}(t-1)\right]_{rl}
\left[\widetilde{\mathbf{S}}_{\mathbf{P}}(t-1)\right]_{rn}
\right]\mathbb{E}\left[\left(\frac{e_{rn}(t)}{q_{rn}}-1\right)\right]
\nonumber
\\
\nonumber
& = & 0
\end{eqnarray}
}
where we have repeatedly used the fact that
\begin{equation}
\label{reply5}
\mathbb{E}\left[\left(\frac{e_{rn}(t)}{q_{rn}}-1\right)\right]=0
\end{equation}
From eqns.~(\ref{reply1}-\ref{reply4}) it is then clear that
\begin{equation}
\label{reply6}
\mathbb{E}\left[\widetilde{\mathbf{S}}^{T}_{\mathbf{P}}(t-1)
\widetilde{\mathbf{P}}(t-1)\right]=0
\end{equation}
In a similar way, it can be shown that the other dropped crossed
terms in eqn.~(\ref{th:conv12}) are zero-mean.

We note that there exist constants,~$k_{3},k_{4},k_{5},k_{6}>0$,
such that
\begin{equation}
\label{th:conv13} \|\mathbf{x}\|^{2}\leq
k_{3}\|\mathbf{x}-\mathbf{d}^{\ast}\|^{2}+k_{4},~~\|\mathbf{x}\|\leq
k_{5}\|\mathbf{x}-\mathbf{d}^{\ast}\|^{2}+k_{6}.
\end{equation}
Hence, from~(\ref{th:conv12}) and~(\ref{th:conv13}), we have
\begin{eqnarray}
\label{th:conv14}
\mathbb{E}\left\|\mathbf{\Gamma}\left(t,\mathbf{x},\omega\right)\right\|^{2} &
\leq & k_{7}\left\|\mathbf{x}-\mathbf{d}^{\ast}\right\|^{2}+k_{8}\nonumber, \\
& \leq & k_{9}\left(1+V\left(\mathbf{x}\right)\right),
\end{eqnarray}
where,~$k_{7},k_{8}>0$ and~$k_{9}=\max\left(k_{7},k_{8}\right)$. Combining
eqns.~(\ref{th:conv11},\ref{th:conv14}) we note that
assumption~\textbf{4)} in Theorem~\ref{RM} is satisfied, as
\begin{equation}
\label{th:conv15}
\left(\mathbf{R}\left(\mathbf{x}\right),
V_{\mathbf{x}}\left(\mathbf{x}\right)\right)\leq
0,\:\forall\mathbf{x}.
\end{equation}
Hence, all the conditions of Theorem~\ref{RM} are satisfied and we
conclude that
\begin{equation}
\label{th:conv16}
\mathbb{\mathbf{P}}\left[\lim_{t\rightarrow\infty}\mathbf{x}^{j}(t)
=\left(\mathbf{I}-\mathbf{P}- \mathbf{S}_{\mathbf{P}}\right)^{-1}
\left(\mathbf{B}+\mathbf{S}_{\mathbf{B}}\right)\mathbf{u}^{j}\right]=1.
\end{equation}
Since,~(\ref{th:conv16}) holds for all~$j$, and~$j$ takes
only a finite number of values ($1\leq j\leq m$), we have
\begin{equation}
\label{th:conv17}
\mathbb{\mathbf{P}}\left[\lim_{t\rightarrow\infty}\mathbf{x}^{j}(t)
=\left(\mathbf{I}-\mathbf{P}-\mathbf{S}_{\mathbf{P}}\right)^{-1}
\left(\mathbf{B}+\mathbf{S}_{\mathbf{B}}\right)\mathbf{u}^{j},\,\,\forall
j\right]=1.
\end{equation}
\end{proof}
We now interpret Theorem~\ref{conv}. Referring to the partitioning
of the~$\mathbf{C}(t)$ matrix in~(\ref{algass:2}), we have
\begin{equation}
\label{conv10} \mathbf{C}(t) = \left[\begin{array}{ll}
                    \mathbf{U}\\
                    \mathbf{X}(t)
                   \end{array}
          \right],
\end{equation}
where each row of~$\mathbf{X}(t)$ corresponds to an estimated
sensor location at time~$t$. Theorem~\ref{conv} then states that,
starting with any initial guess,
~$\mathbf{X}(0)\in\mathbb{R}^{M\times m}$, of the unknown sensor
locations, the state sequence,~$\{\mathbf{X}(t)\}_{t\geq 0}$,
generated by the DLRE algorithm converges a.s., i.e.,
\begin{equation}
\label{conv11}
\mathbb{\mathbf{P}}\left[\lim_{t\rightarrow\infty}\mathbf{X}(t)
=\left(\mathbf{I}-\mathbf{P}- \mathbf{S}_{\mathbf{P}}\right)^{-1}
\left(\mathbf{B}+\mathbf{S}_{\mathbf{B}}\right)\mathbf{U}\right]=1.
\end{equation}
However, it follows from Lemma~\ref{lemD}, that the exact
locations of the unknown sensors are given by
\begin{equation}
\label{conv12} \mathbf{X}^{\ast} =
\left(\mathbf{I}-\mathbf{P}\right)^{-1}\mathbf{B}\mathbf{U}.
\end{equation}
Thus, the steady state estimate given by the DLRE algorithm is not
exact, and, to characterize its performance, we introduce the
following notion of localization error,~$e_{l}$, as
\begin{equation}
\label{conv13}
e_{l}=\left\|\left(\mathbf{I}-\mathbf{P}
-\mathbf{S}_{\mathbf{P}}\right)^{-1}
\left(\mathbf{B}+\mathbf{S}_{\mathbf{B}}\right)\mathbf{U}
-\left(\mathbf{I}-\mathbf{P}\right)^{-1}\mathbf{B}\mathbf{U}\right\|.
\end{equation}
We note that~$e_{l}$ is only a function of
~$\mathbf{S}_{\mathbf{P}}, \mathbf{S}_{\mathbf{B}}$, the non-zero biases in the system matrix computations, resulting from noisy inter-sensor distance measurements (see, Section~\ref{AlgAss}.) We note that the DLRE algorithm is robust to random link failures and additive channel noise in inter-sensor communication. In fact, it is also robust to the zero-mean random errors in the system matrix computations, and only affected by the fixed non-zero biases. Note
that~$e_{l}=0$ for~$\mathbf{S_P}=\mathbf{S_B}=0$. Clearly, if we assume sufficient accuracy in the inter-sensor distance computation process, so that the biases,~$\mathbf{S}_{\mathbf{P}}, \mathbf{S}_{\mathbf{B}}$, are small, the steady state error~$e_{ss}$ will also be negligible, even in a random sensing environment. These are illustrated by numerical studies provided in Section~\ref{num}.

\section{Numerical Studies}\label{num}
We divide the numerical study of DILOC into the following parts. First, we present DILOC in the deterministic case, i.e., we have no communication noise, no link failures, and the required inter-sensor distances are known precisely. Second, we present DILOC when there is communication noise and link failures. Third, we consider noise on the distance measurements that results in random system matrices,~$\mathbf{\widehat{P}}(t)$ and~$\mathbf{\widehat{B}}(t)$, as given in~\eqref{algass:4}--\eqref{algass:6}; we study both the biased~$\left(\mathbf{S}_\mathbf{P}\neq 0, \,\mathbf{S}_\mathbf{B}\neq 0\right)$ and unbiased~$\left(\mathbf{S}_\mathbf{P}=\mathbf{S}_\mathbf{B}= 0\right)$ cases  in the following. In the end, we present studies of DILOC in the presence of all random scenarios.

\begin{figure}
\centering
\subfigure[]
{
    \label{coords_det}
    \includegraphics[width=1.2in]{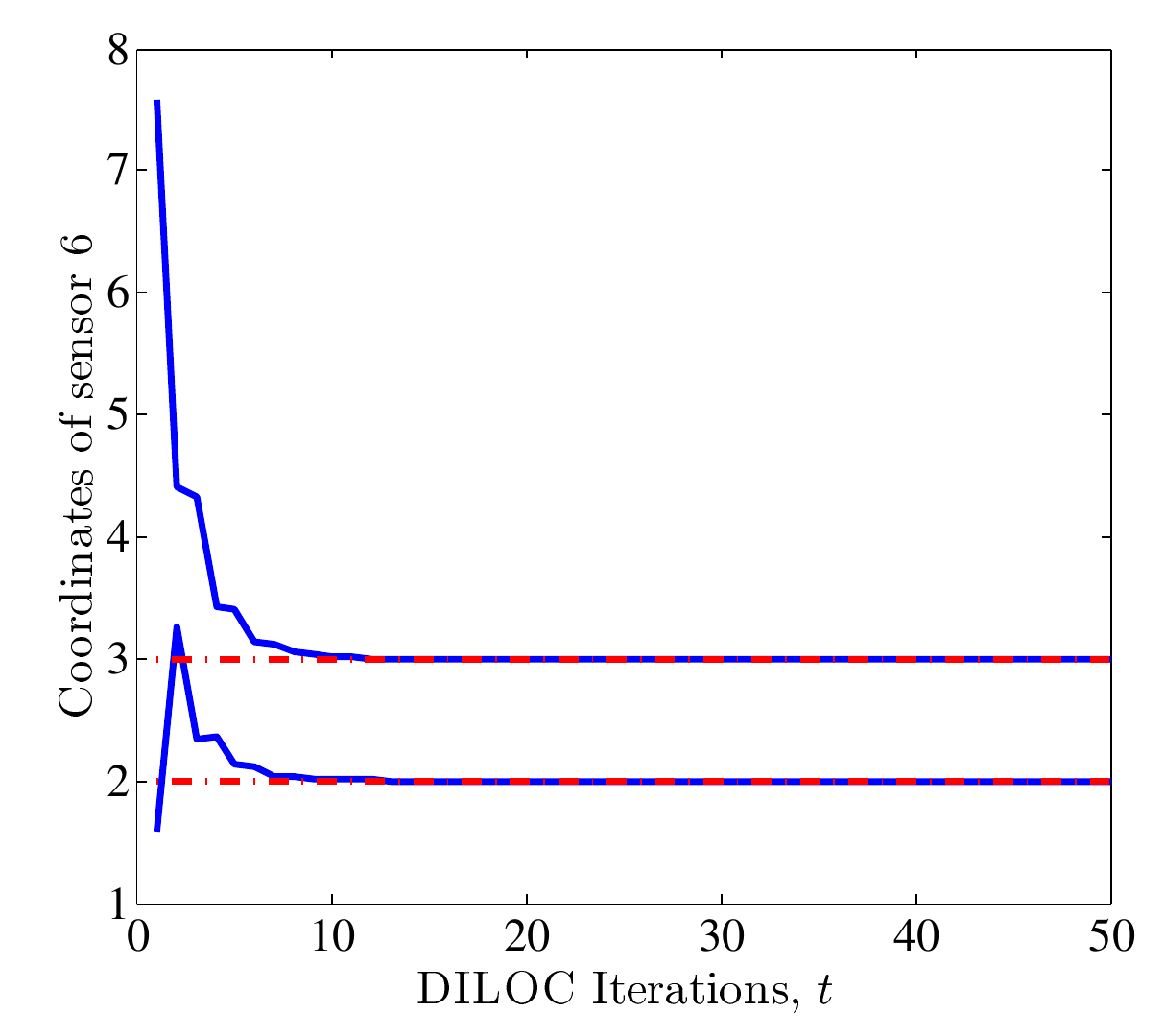}
}
\hspace{.1cm}
\subfigure[]
{
    \label{det_traj}
    \includegraphics[width=1.5in]{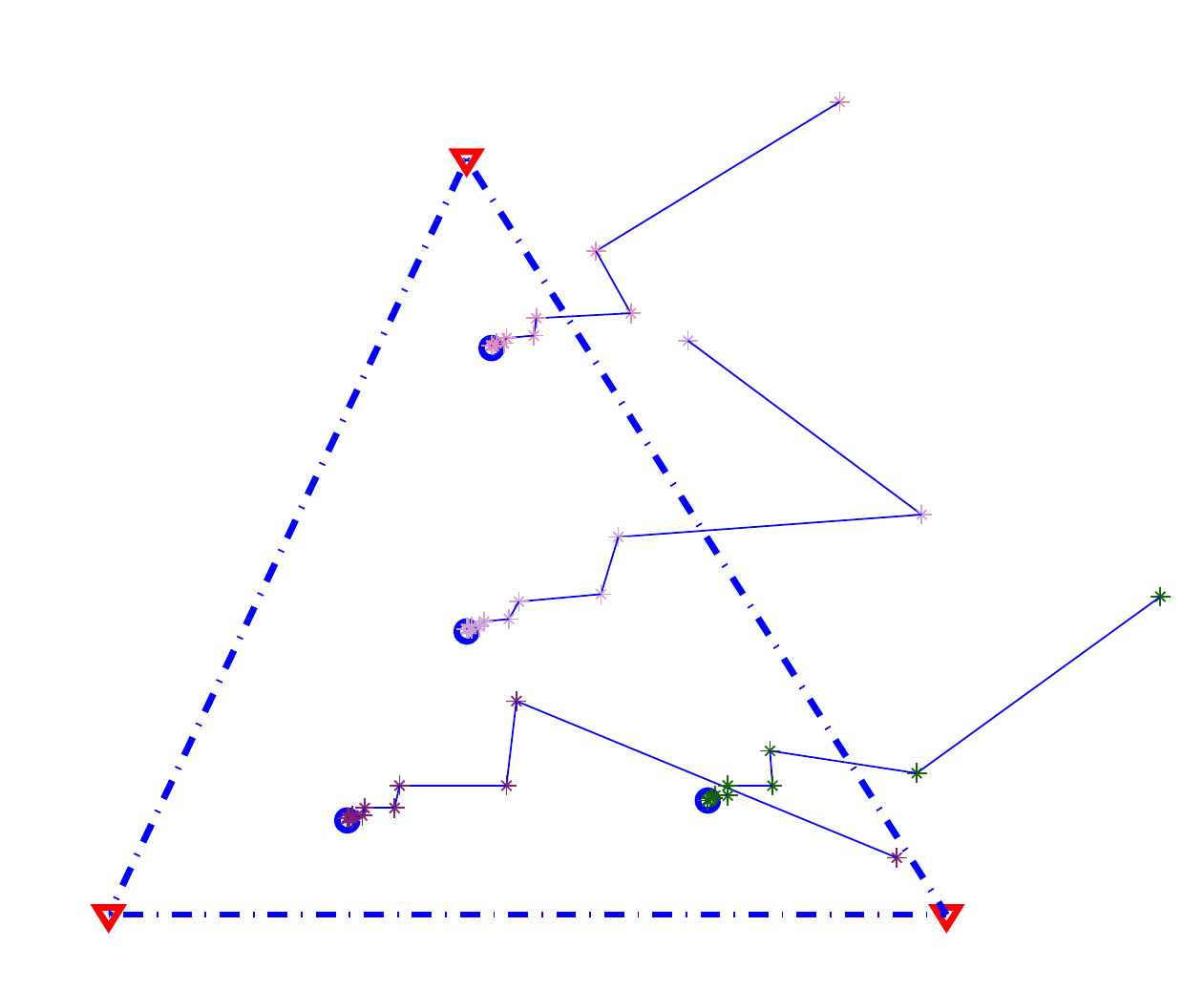}
}
\hspace{.1cm}
\subfigure[]
{
    \label{large_tri}
    \includegraphics[width=1.4in]{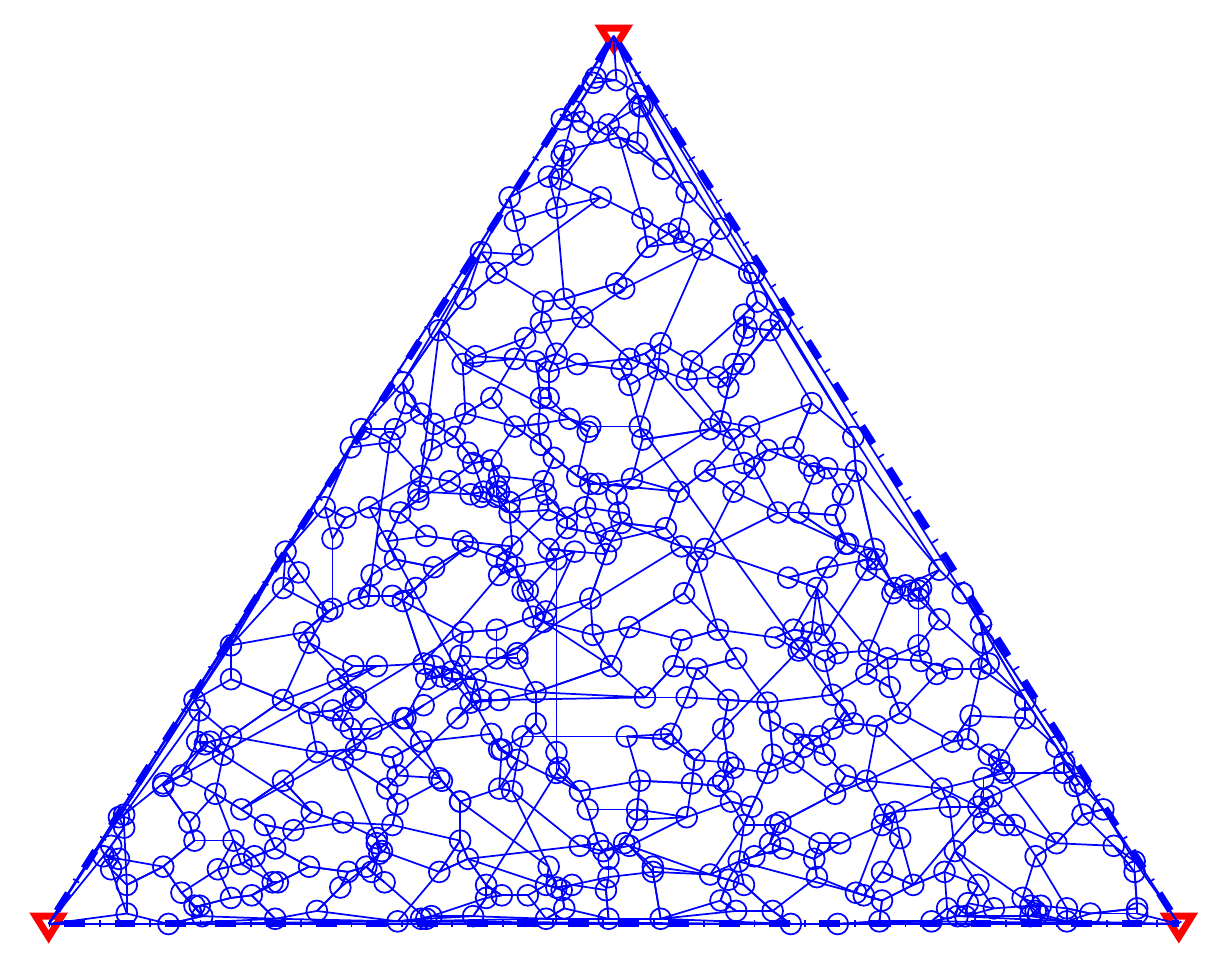}
}
\hspace{.1cm}
\subfigure[]
{
    \label{large_traj}
    \includegraphics[width=1.4in]{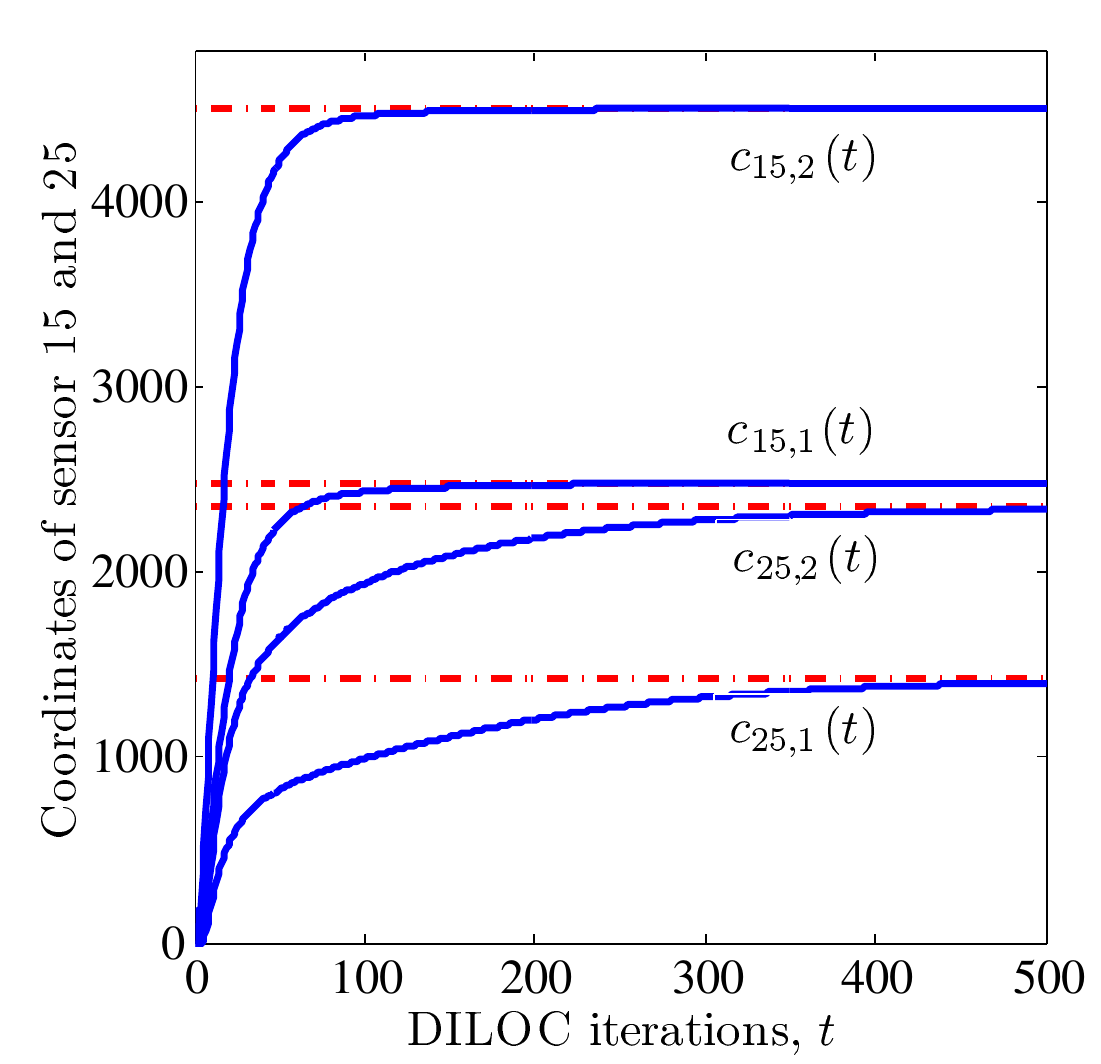}
}
\caption{Deterministic environments: (a) and (b) DILOC algorithm implemented on the example in Section~\ref{example}. (c) An~$N=500$ node network and the respective triangulation sets. (d) DILOC implemented on the network in (c), where the iterations are shown for two arbitrarily chosen sensors.}
\label{saa} 
\end{figure}

{\bf DILOC Algorithm in Deterministic Environments: }We consider the example presented in section~\ref{example}. We have~$N=7$ nodes in~$m=2$-dimensional space, where~$m+1=3$ are the anchors and~$M=4$ are the sensors. DILOC, as given in~\eqref{ex_eq}, is implemented, and the results are shown in Fig.~\ref{coords_det}--\ref{det_traj}. Fig.~\ref{coords_det} shows the estimated coordinates of sensor~$6$ and Fig.~\ref{det_traj} shows the trajectories of the estimated coordinates for all the sensors with random initial condition. Fig.~\ref{large_tri} and Fig.~\ref{large_traj} show DILOC for a network of~$N = 500$ nodes.

\begin{figure}
\centering
\subfigure[]
{
    \label{lf_cn_examp_coords}
    \includegraphics[width=2in]{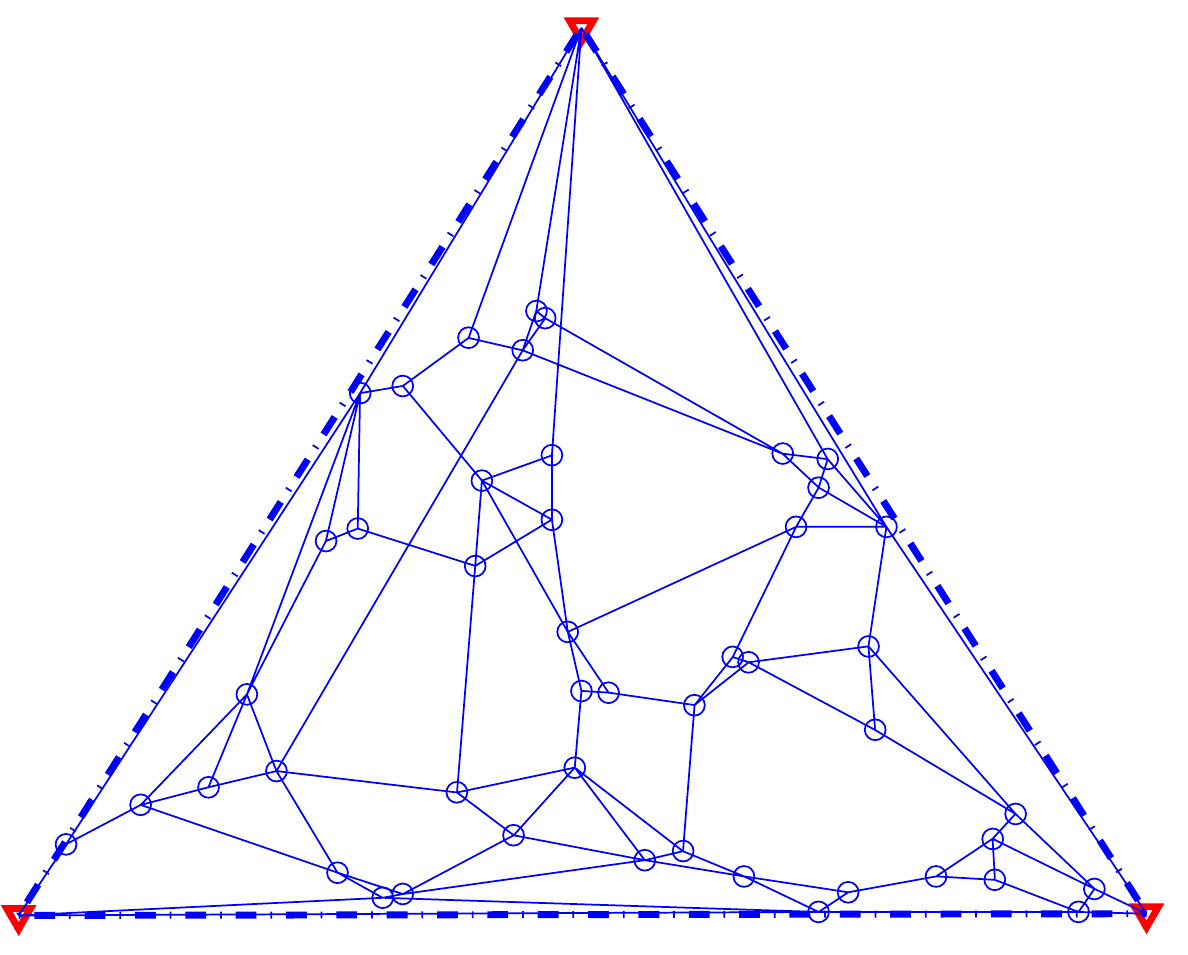}
}
\hspace{1cm}
\subfigure[]
{
    \label{lf_cn_examp_tri}
    \includegraphics[width=2in]{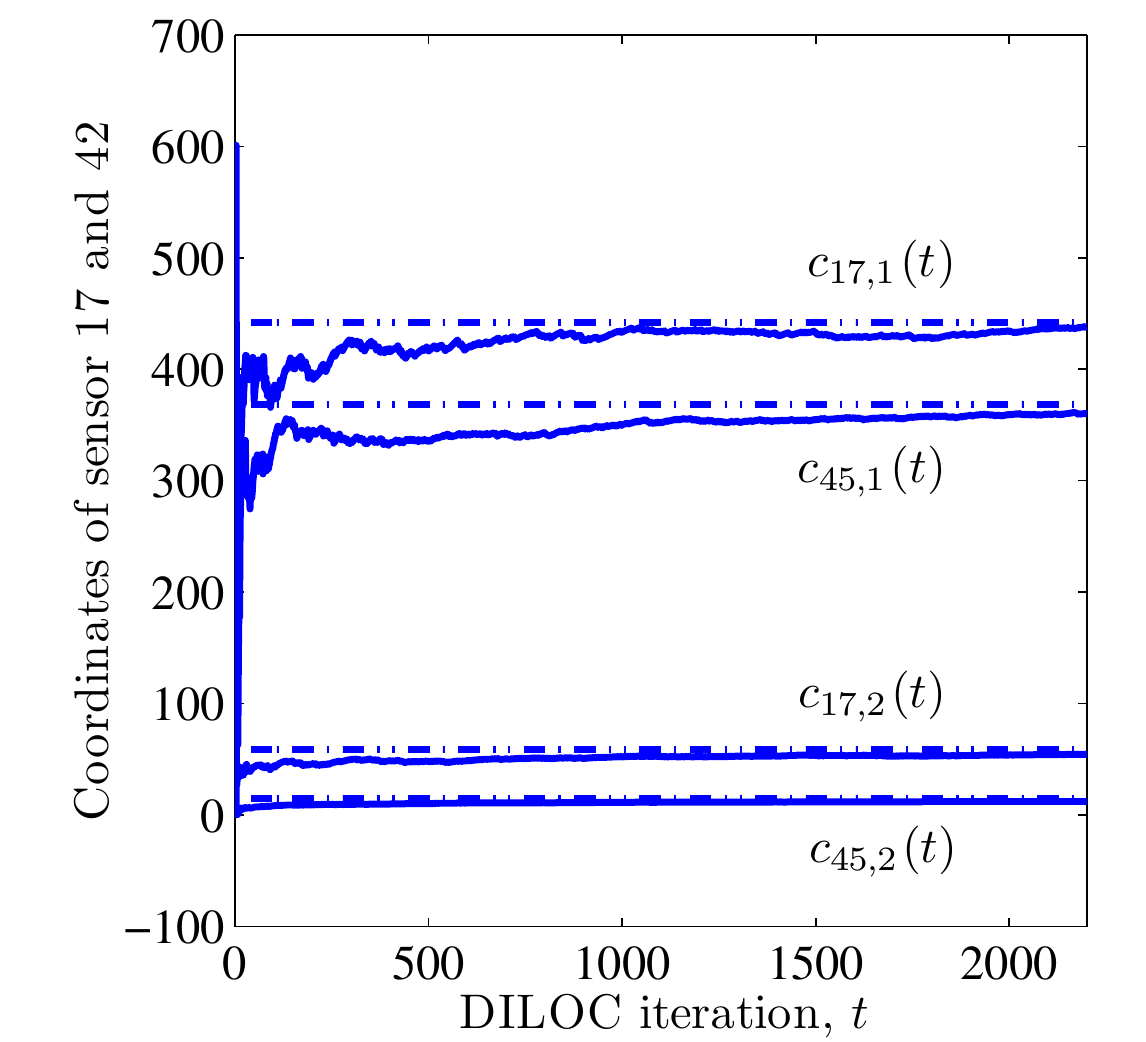}
}
\caption{Effect of communication noise and link failures: (a) An~$N=50$ node network and the respective triangulation sets. (d) DLRE (with a decreasing weight sequence,~$\alpha=\frac{4}{t+1}$) implemented on the network in (a), where the iterations are shown for two arbitrarily chosen sensors.}
\label{sa} 
\end{figure}

{\bf DILOC Algorithm with Communication Noise and Link Failures: }We consider the same example of section~\ref{example}, but, include noise in the communication and link failures. All the communication links are active~$90\%$ of the time, i.e.,~$q_{ln}=0.9,~\forall~n~\mbox{s.t.}\,n\in\Theta_l$, as discussed in {\bf (C2)}, and include an additive communication noise that is Gaussian i.i.d.~with zero-mean and variance~$1/M$ (roughly speaking, this is equivalent to having a unity variance for the entire network). In this scenario, we employ DILOC with a decreasing weight sequence,~$\alpha(t)=\frac{4}{t+1}$ and the results are presented in Fig.~\ref{lf_cn_examp_coords}and Fig.~\ref{lf_cn_examp_tri}.

\begin{figure}
\centering
\subfigure[]
{
    \label{DILOC_dn0}
    \includegraphics[width=1.7in]{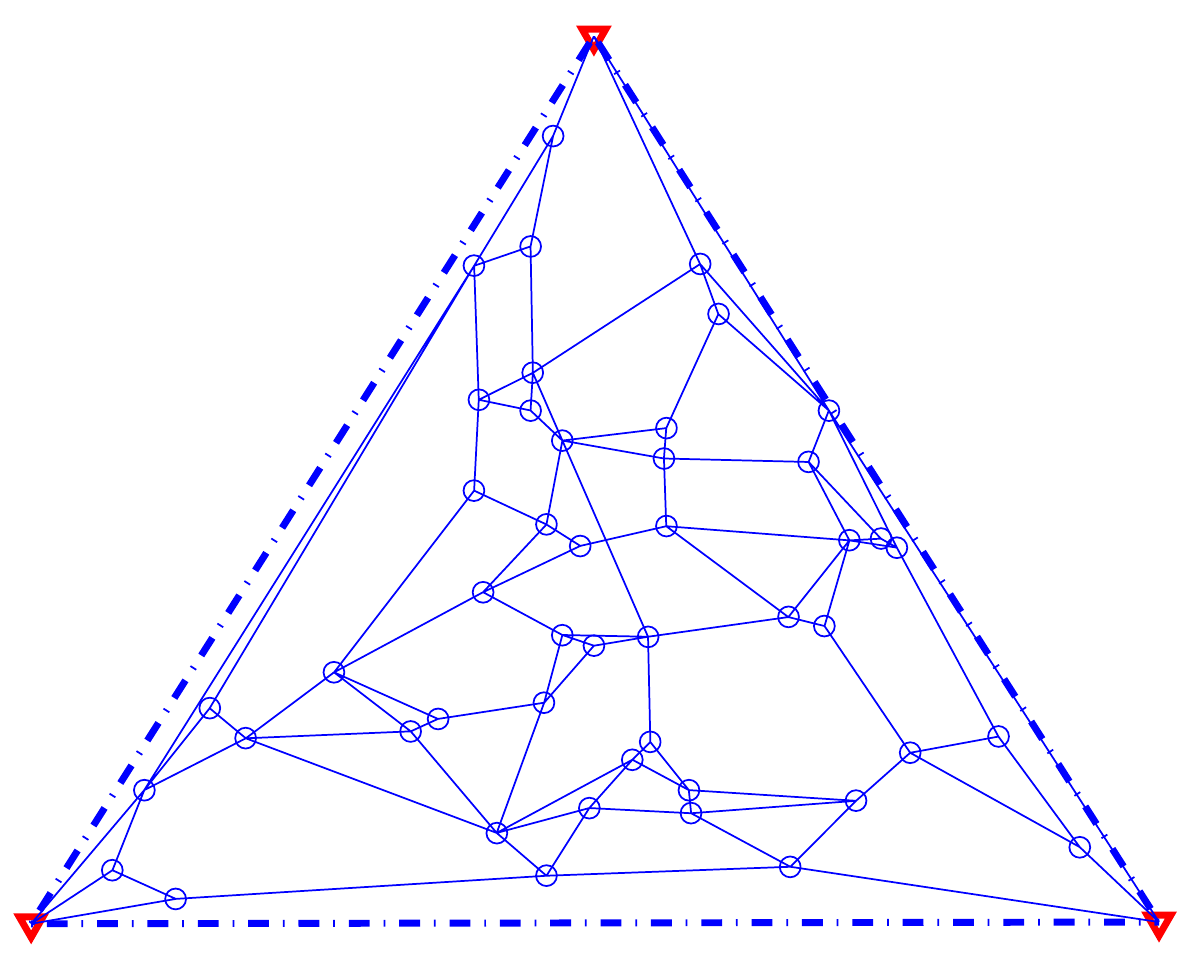}
}
\hspace{.5cm}
\subfigure[]
{
    \label{DILOC_dn1}
    \includegraphics[width=1.5in]{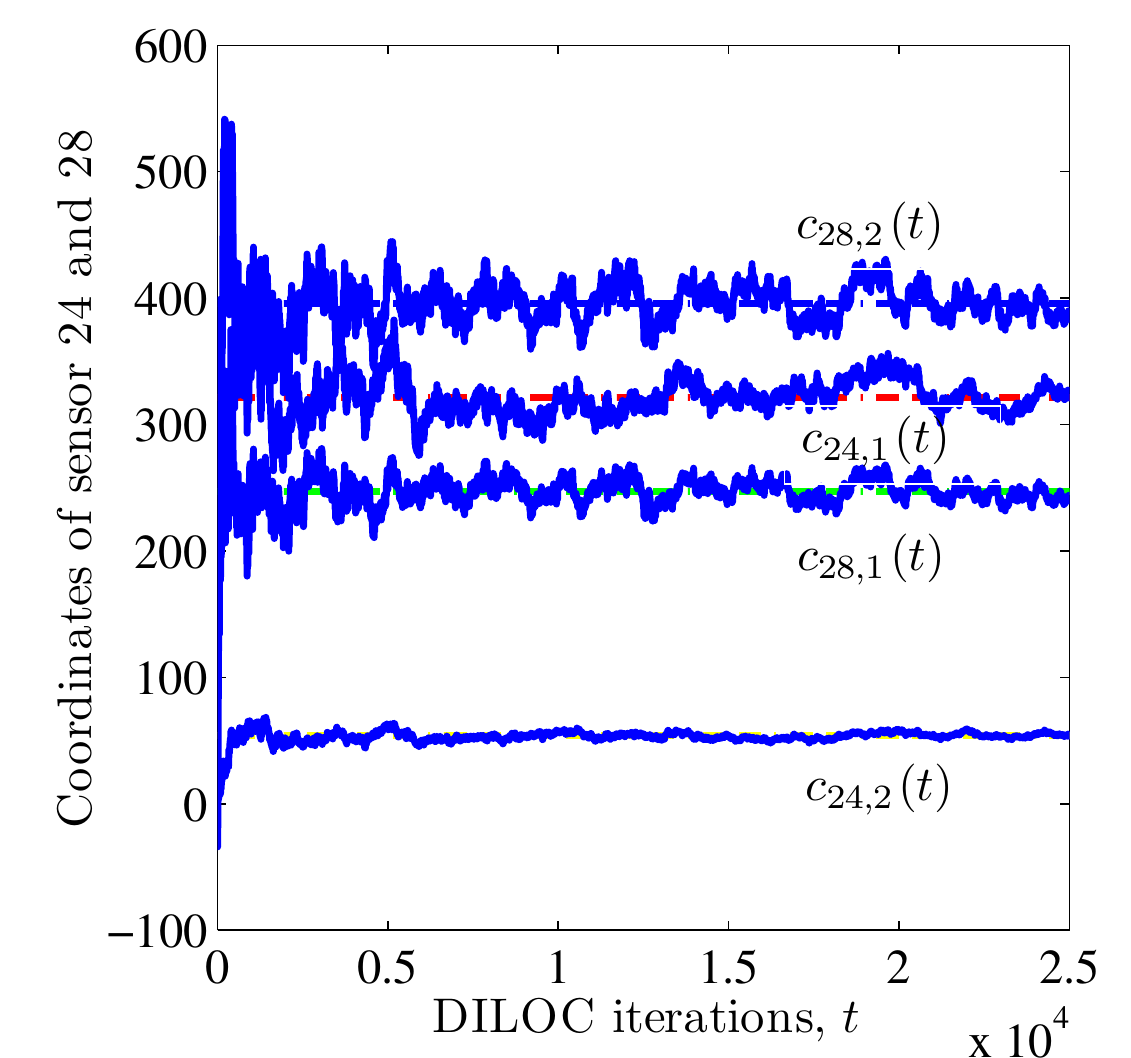}
}

\caption{Effect of noisy distance measurements: (a) An~$N=50$ node network and the respective triangulation sets. (d) DLRE (with a decreasing weight sequence,~$\alpha=\frac{1}{t^0.55}$) implemented on the network in (a), where the iterations are shown for two arbitrarily chosen sensors.}
\label{suss} 
\end{figure}

\begin{figure}
\centering
\subfigure[]
{
    \label{DILOC_bs1}
    \includegraphics[width=1.7in]{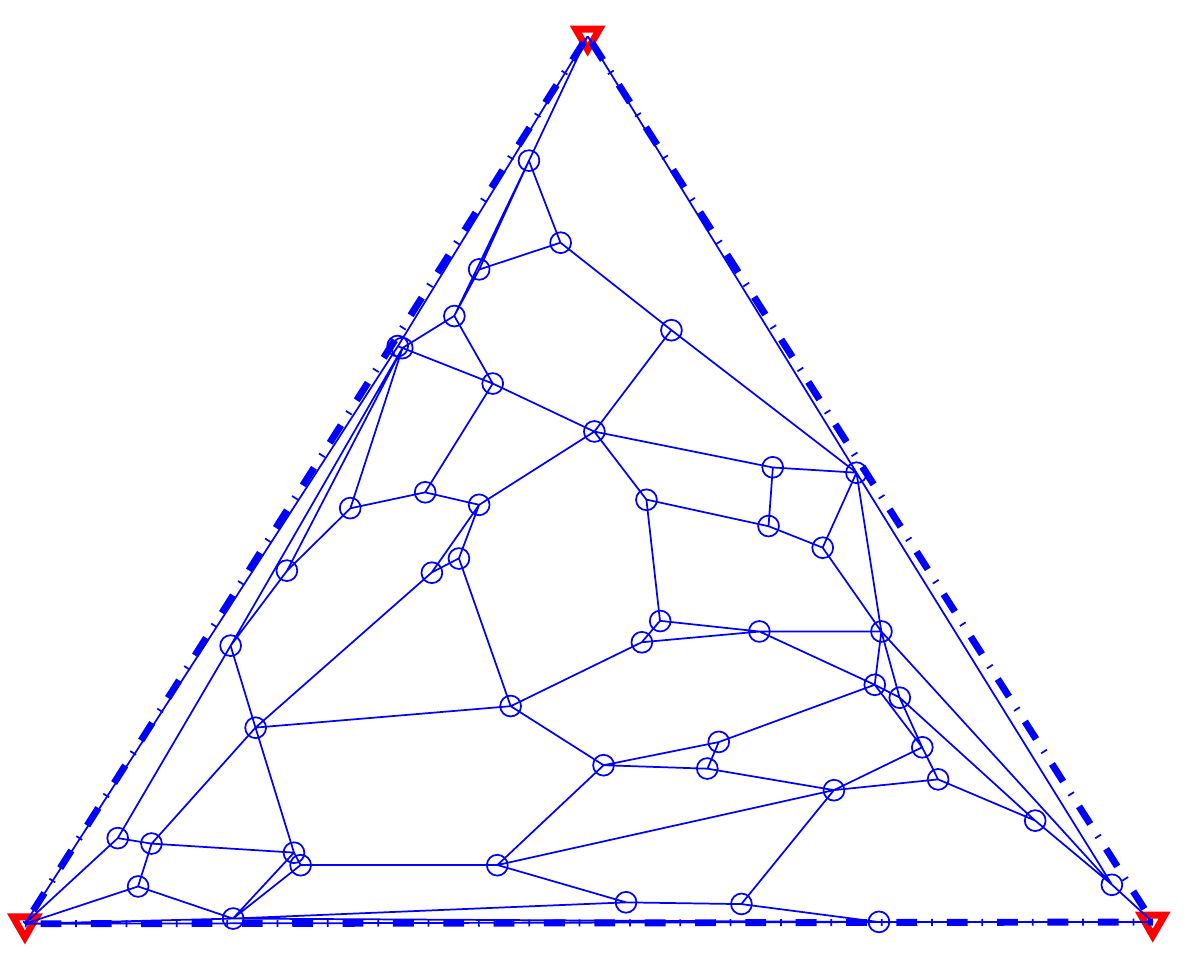}
}
\hspace{.5cm}
\subfigure[]
{
    \label{DILOC_bs2}
    \includegraphics[width=1.5in]{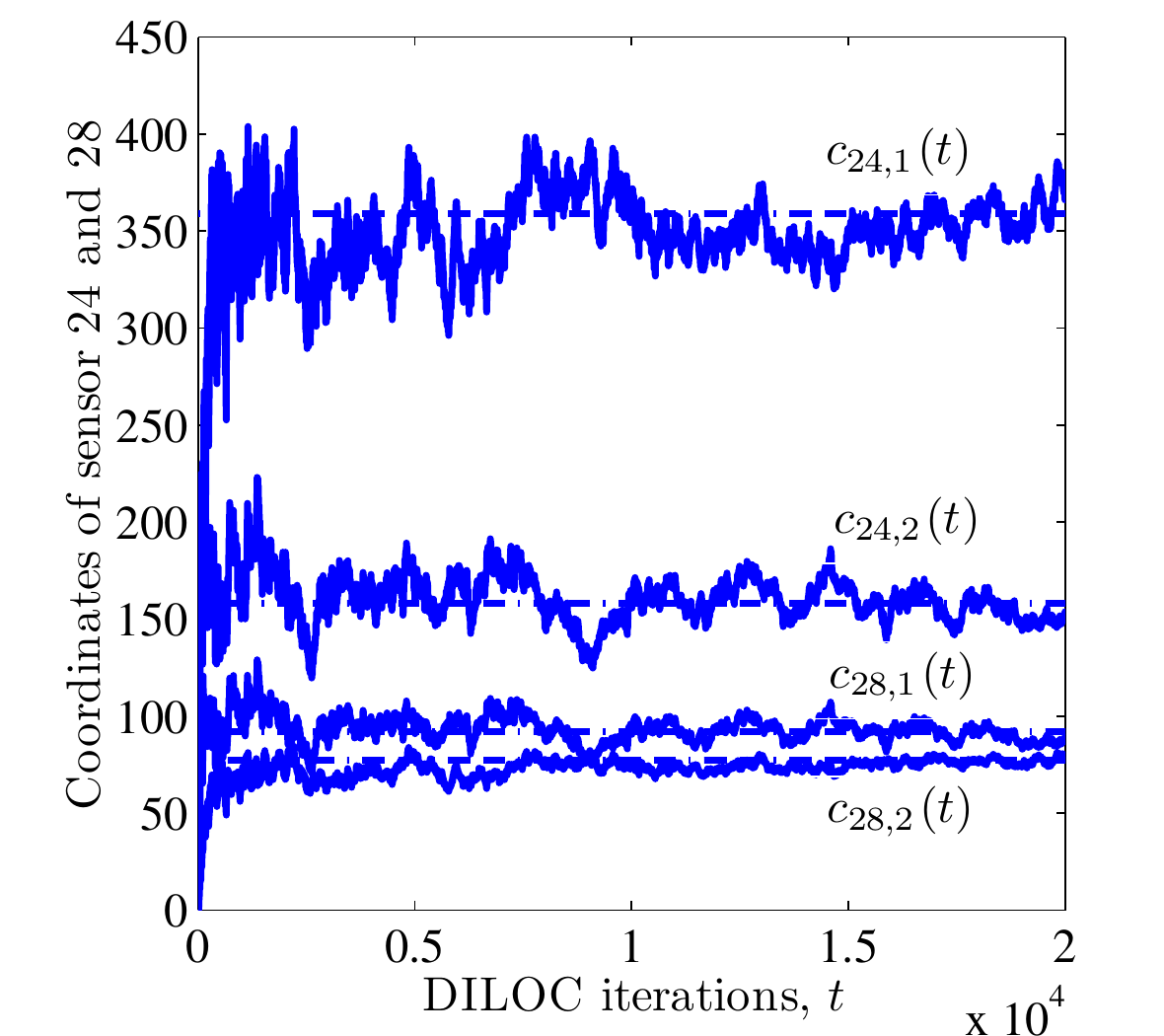}
}

\caption{Random environments (Noisy distances, communication noise, link failures): (a) An~$N=50$ node network and the respective triangulation sets. (d) DLRE (with a decreasing weight sequence,~$\alpha=\frac{1}{t^0.55}$) implemented on the network in (a), where the iterations are shown for two arbitrarily chosen sensors.}
\label{suss} 
\end{figure}

{\bf DILOC with Noisy Distance measurements: }We now consider noise on the distance measurements. We assume that we have a reasonable estimate of the required distances such that it translates into a small perturbation of the system matrices,~$\mathbf{\widehat{B}}(t)$ and~$\mathbf{\widehat{P}}(t)$. The matrices~$\mathbf{\widetilde{S}}_\mathbf{P},\mathbf{\widetilde{S}}_\mathbf{B}$ are zero-mean Gaussian i.i.d.~perturbations with variance~$0.1$ (small signal perturbation, note that the non-zero elements of both the matrices,~$\mathbf{B}$ and~$\mathbf{P}$ lie in the range of~$0$ and~$1$). For a network of~$N=50$ nodes in Fig.~\ref{DILOC_dn0}, we implement DLRE, with a decreasing weight sequence,~$\alpha=\frac{1}{t^0.55}$, in Fig.~\ref{DILOC_dn1}. Finally, Fig.~\ref{DILOC_bs1} shows a network of~$N=50$ nodes, where DLRE, with a decreasing weight sequence,~$\alpha=\frac{1}{t^0.55}$, with all of the above random scenarios is implemented in Fig.~\ref{DILOC_bs2}.

\section{Conclusions}\label{conc}
The paper studies a distributed iterative sensor localization algorithm in
~$m-$dimensional Euclidean space,~$\mathbb{R}^m~(m\geq1)$, that
finds the location coordinates of the sensors in a sensor network with only local communication. The algorithm uses the minimal number,~$m+1$, of anchors (sensors with known location) to
localize an arbitrary number,~$M$, of sensors that lie in the
convex hull of these~$m+1$ anchors. In the deterministic case, i.e., when no noise affects the inter-sensor communication, the inter-sensor distances are known with no errors, and the communication links do not fail, we show that our distributed algorithms, DILOC and DILOC-REL, lead to convergence to the exact sensor locations. For the random environment scenario, where inter-sensor communication links may fail randomly, transmitted data is distorted by noise, and inter-sensor distance information is imprecise, we show that our modified algorithm, DLRE, leads to almost sure convergence of the iterative location estimates, and in this case we explicitly characterize the resulting error between the exact sensor locations and the converged estimates. Numerical simulations illustrate the behavior of the algorithms under different field conditions.

\appendices
\section{Convex Hull Inclusion test}\label{CHIT}
\begin{figure}
\centering
\subfigure[]
{
    \label{CHIT1.pdf}
    \includegraphics[width=2in]{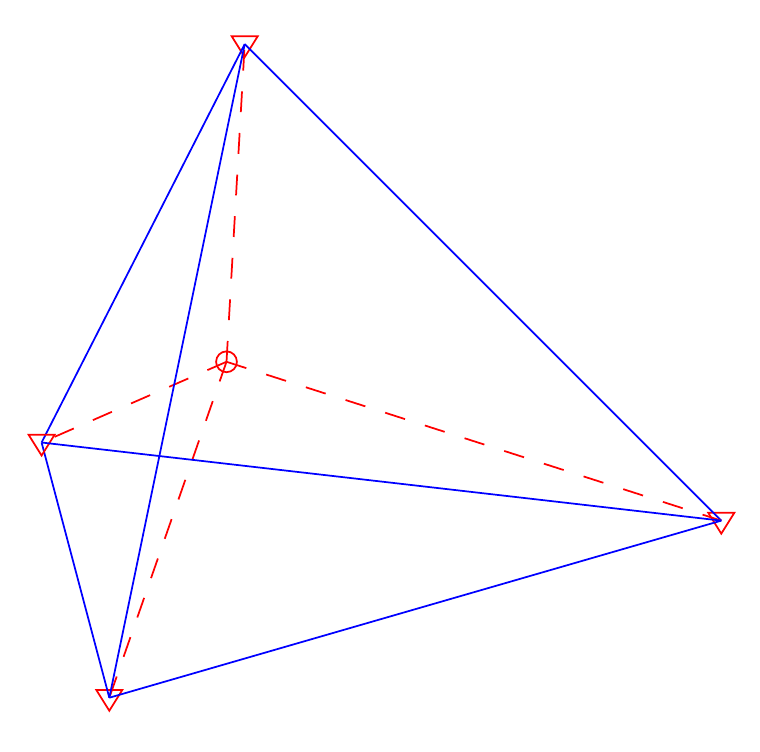}
}
\hspace{.5cm}
\subfigure[]
{
    \label{CHIT2.pdf}
    \includegraphics[width=2in]{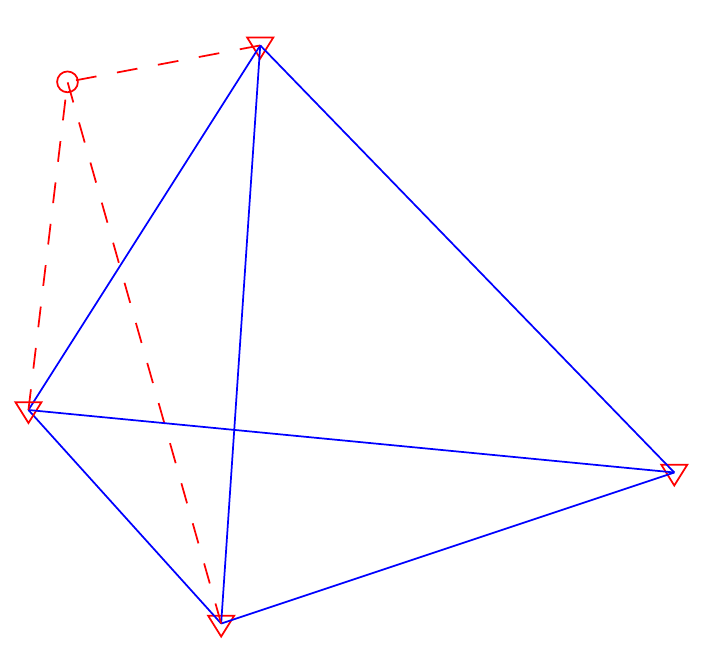}
}
\caption{Convex Hull Inclusion Test (m=3): The sensor~$l$ is shown by a `$\circ$', whereas, the anchors in~$\kappa$ are shown by~`$\nabla$'. (a)~$l\in\mathcal{C}(\kappa)\Rightarrow A_{\kappa} = A_{\kappa\cup\{l\}}$, (b)~$l\notin\mathcal{C}(\kappa)\Rightarrow A_{\kappa} < A_{\kappa\cup\{l\}}$.}
\label{CHIT_fig} 
\end{figure}

We now give an algorithm that tests if a given sensor,~$l\in\mathbb{R}^m$, lies in the convex hull of~$m+1$ nodes in a set,~$\kappa$, using only the mutual distance information among these~$m+2$ nodes ($\kappa\cup\{l\}$). Let~$\kappa$ denote the set of~$m+1$ nodes and let~$\mathcal{C}(\kappa)$ denote the convex hull formed by the nodes in~$\kappa$. Clearly, if~$l\in\mathcal{C}(\kappa)$, then the convex hull formed by the nodes in~$\kappa$ is the same as the convex hull formed by the nodes in~$\kappa\cup\{l\}$, i.e.,
\begin{equation}
\mathcal{C}(\kappa) = \mathcal{C}(\kappa\cup\{l\}),\qquad\mbox{if }l\in\mathcal{C}(\kappa).
\end{equation}
With the above equation, we can see that, if~$l\in\mathcal{C}(\kappa)$, then the generalized volumes of the two convex sets,~$\mathcal{C}(\kappa)$ and~$\mathcal{C}(\kappa\cup\{l\})$, should be equal. Let~$A_{\kappa}$ denote the generalized volume of~$\mathcal{C}(\kappa)$ and let~$A_{\kappa\cup\{l\}}$ denote the generalized volume of~$\mathcal{C}(\kappa\cup\{l\})$, we have
\begin{eqnarray}
A_{\kappa} &=& A_{\kappa\cup\{l\}}, \nonumber\\&=& \sum_{k\in\kappa}A_{\kappa\cup\{l\}\setminus\{k\}}, \qquad\mbox{if }l\in\mathcal{C}(\kappa).
\end{eqnarray}
Hence, the test becomes\begin{eqnarray}l\in\mathcal{C}(\kappa),\qquad \mbox{if}\,\, \sum_{k\in\kappa}A_{\kappa\cup\{l\}\setminus\{k\}} = A_{\kappa},\\ l\notin\mathcal{C}(\kappa),\qquad \mbox{if}\,\, \sum_{k\in\kappa}A_{\kappa\cup\{l\}\setminus\{k\}} > A_{\kappa}.\end{eqnarray} This is also shown in Figure~\ref{CHIT_fig}. The above inclusion test is based entirely on the generalized volumes, which can be calculated using only the distance information in the Cayley-Menger determinants.

\section{Cayley-Menger Determinant}\label{CMdet}
Let~$\kappa$ be a set of~$m+1$ points (sensors) in~$\mathbb{R}^m$, and~$d_{lj}$ be the inter-sensor distance between~$l$ and~$j$. The
generalized volume,~$A_\kappa$, of the convex hull
of the points in~$\kappa$ can be computed by the Cayley-Menger
determinant, see, e.g.,~\cite{cayley_men:86}. The Cayley-Menger
determinant is the determinant of an~$m+2\times m+2$ (symmetric) matrix that
relates to the generalized volume,~$A_\kappa$, of the convex hull,
~$\mathcal{C}(\kappa)$, of the~$m+1$ points in~$\mathbb{R}^m$
through an integer sequence,~$s_{m+1}$. The Cayley-Menger
determinant is given by \begin{equation} s_{m+1}
A_\kappa^2=\left|\begin{array}{ccccccc}
0 & 1 & 1 & 1 &  & \ldots & 1\\
1 & 0 & d_{12}^2 & d_{13}^2 & & \ldots & d_{1,m+1}^2\\
1 & d_{21}^2 & 0 & d_{23}^2 & d_{24}^2 & \ldots & d_{2,m+1}^2\\
1 & d_{31}^2 & d_{32}^2 & 0 & d_{34}^2 & \ddots & \vdots \\
&&d_{42}^2&d_{43}^2&0&\ddots&d_{m-1,m+1}^2\\
\vdots&\vdots&\vdots&\ddots&\ddots&\ddots&d_{m,m+1}^2\\
1&d_{m+1,1}^2&d_{m+1,1}^2&\ldots&d_{m+1,m-1}^2&d_{m+1,m}^2& 0
\end{array}\right|,
\end{equation}
where
 \begin{equation}s_{m}= {(-1)^{m+1}}{2^m(m!)^2},\qquad\qquad m=\{0,1,2,\ldots\},
 \end{equation}
 and its first few terms are~$-1, 2, -16, 288, -9216, 460800, \ldots.$
\section{Important Results}\label{IR}
\begin{lem}\label{lem2} If the matrix,~$\mathbf{P}$, corresponds to the transition probability matrix associated to the transient states of an absorbing Markov chain, then
\begin{eqnarray}\label{lem2_eq}\lim_{t\rightarrow\infty}\mathbf{\mathbf{P}}^{t+1} &=& \mathbf{0}.\end{eqnarray}
\end{lem}
\begin{proof}
For such a matrix,~$\mathbf{P}$, we have
  \begin{equation}
  \label{lemmm}
  \rho(\mathbf{P})<1,
  \end{equation}
from Lemma~8.3.20 and Theorem~8.3.21 in \cite{plemmons:79}, where~$\rho(\cdot)$ denotes the spectral norm of a matrix and~\eqref{lem2_eq} follows from~\eqref{lemmm}.
\end{proof}
\begin{lem}\label{lem1}If the matrix,~$\mathbf{P}$, corresponds to the transition probability matrix associated to the transient states of an absorbing Markov chain, then
\begin{eqnarray}\lim_{t\rightarrow\infty}\sum_{k=0}^{t+1}\mathbf{\mathbf{P}}^{k}&=&\left(\mathbf{\mathbf{I}-\mathbf{P}}\right)^{-1}.\end{eqnarray}
\end{lem}
\begin{proof}

The proof follows from Lemma~\ref{lem2} and Lemma~6.2.1 in~\cite{plemmons:79}.
\end{proof}

\bibliographystyle{IEEEbib}
\bibliography{ref}

\end{document}